\documentstyle[preprint,aps,eqsecnum,amssymb]{revtex}
\def\pacs#1{\par %
\bgroup
\hsize\columnwidth \parindent0pt
\if@twocolumn\else\leftskip=0.10753\textwidth \rightskip\leftskip\fi
\ifdim\prevdepth=-1000pt \prevdepth0pt\fi
\dimen0=-\prevdepth \advance\dimen0 by20pt\nointerlineskip
\vbox to28pt{\small\vrule height\dimen0 width0pt\relax\ifdraft#1\fi\vfill}%
\egroup
\if@twocolumn\vskip1pc\fi
\ifpreprintsty
\penalty10000\vfill
\newpage
\fi
}
\begin{document}
\draft
\title{Computation of Lie Transformations from a Power Series: Bounds and 
Optimum Truncation}
\author{Ivan Gjaja}
\address{Center for Nonlinear Studies, Los Alamos National Laboratory,
Los Alamos, NM 87545}
\maketitle
\begin{abstract}
The problem considered is the computation of an infinite product (composition) 
of Lie transformations generated by homogeneous polynomials of increasing order
from a given convergent power series. Bounds are computed for the
infinitesimal form of Lie transformations. The results obtained do not
guarantee convergence of the product. Instead, the optimum 
truncation is determined by minimizing the terms of order $n+1$ that 
remain after the first $n$ Lie transformations have been applied.

\vskip 3.6truein
\noindent LA--UR--94--3671

\end{abstract}

\section{Introduction}
\label{sec:1}

A method based on an infinite product (composition) of Lie transformations
(exponentiated vector fields) generated by homogeneous polynomials of
increasing order was developed a long time ago in order to efficiently perform
perturbative calculations in Hamiltonian systems when the small parameters
are the dynamical variables themselves \cite{df}. In particular, it was
shown in Ref.~\cite{df} that the product can be computed from, or be used
to compute, a power series in the dynamical variables. The relation between
Lie transformations and power series, however, was established only at a 
formal level, that is order by order.  

More recent work has provided firm bounds on the results that can be 
obtained using the method. In Refs.~\cite{fassben,fasso,koseleff},
for example, a variant of the method is applied to the problem of 
bringing a Hamiltonian function or a Hamiltonian vector field to normal form.
In Ref.~\cite{me}, and in Ref.~\cite{koseleff} 
for Hamiltonian systems, sufficient conditions are given on the
coefficients of the polynomials and on the domain of the dynamical variables
such that the infinite product of Lie transformations is convergent. 
In this paper we turn to the 
construction of Lie transformations from a power series. Assuming that the 
power series has
a finite domain of absolute convergence, we obtain bounds for the norms of 
the vector fields that are computed from such a series and develop a procedure
for determining the optimum truncation of the product of Lie transformations.
(Note that the analogous problem with power series for Refs.~\cite{fassben}
and \cite{fasso} is the one in which only the first nonlinear term in the
series is nonvanishing.)
As in Ref.~\cite{me}, we do not require that Lie transformations 
be symplectic (i.e. that they arise from Hamiltonian systems); rather,
the vector fields are taken to be arbitrary homogeneous polynomials.

In Section \ref{sec:2} we introduce notation, which for quantities
that appear in both is the same as in Ref.~\cite{me}, and write down an
expression for the coefficients of the polynomials in terms of the 
coefficients of the power series. Section \ref{sec:3} contains two lemmas 
which allow us to pass from the expression for the coefficients 
to an inequality in the form of a recursion relation for
the norms of the vector fields. In Section \ref{sec:4} we then use the 
recursion relation to obtain a bound on the norms, which is the main 
result of the paper. We also provide there an asymptotic expression for the 
bound valid in the limit $n\rightarrow\infty$, where $n$ is the order of
the polynomial. The question of optimum truncation of the product is
considered in Section \ref{sec:5}, and a summary of the results is
given in Section \ref{sec:6}.

\section{Preliminaries}
\label{sec:2}

We work with the transformation $\cal M$ formally defined by
\begin{equation}
{\cal M}z=e^{L_2(z)}e^{L_3(z)}\ldots e^{L_n(z)}\ldots z,\;\;\;\;\;\;
z\in{\Bbb C}^d.
\label{eq:2.1}
\end{equation}
Here  $L_n$ is a vector field
\begin{equation}
L_n(z)=\sum_{j=1}^{d}g_j^{(n)}(z){\partial\over{\partial z_j}},
\label{eq:2.2}
\end{equation}
and $g_j^{(n)}$ a homogeneous polynomial in $z$ of order $n$,
\begin{equation}
g_j^{(n)}(z)=\sum_{|r|=n}
a_{r j}^{(n)} z^r.
\label{eq:2.3}
\end{equation}
The subscript $r$ stands for the collection of indices
$r_1,\dots, r_{d}$, $\;|r|\,{\buildrel\rm def\over =}\,r_1+\dots+r_{d}$,
and $z^r\,{\buildrel\rm def\over =}\,z_1^{r_1}\dots z_{d}^{r_{d}}$.
The exponential of $L_n$ is given by the usual infinite series
\begin{equation}
e^{L_n(z)}=\sum_{s=0}^\infty {1\over{s!}}[L_n(z)]^s,
\label{eq:2.4}
\end{equation}
where $s=0$ corresponds to the identity transformation. 
In the definition of $\cal M$ the linear transformation has been set equal 
to the identity, as its computation is not germane to the problem at hand.
The designation of Eq. (\ref{eq:2.1}) as formal, on the other hand, reflects
the fact that we have not specified a domain in ${\Bbb C}^d$, if such one 
exists, on which the infinite series of Eq. (\ref{eq:2.4}) are convergent for
$n=2,3,\dots$. We also define ${\cal M}_n$
as the product of Lie transformations of the form (\ref{eq:2.1}) truncated
at order $n$.

The properties of Lie transformations and their use in perturbation 
calculations are not discussed further in this paper. The interested reader 
is instead directed
to Refs.\cite{lt} for a sampling of the surveys of the subject.

Suppose we are given the power series
\begin{equation}
P_k(z)=z_k+\sum_{i=2}^\infty \sum_{|r|=i} b^{(i)}_{r k} 
z^r
\label{eq:2.5}
\end{equation}
which has 
a nonvanishing domain of absolute convergence, denoted here by ${\cal D}$. 
(${\cal D}$ evidently includes the origin.) We are going to examine the 
construction of vector 
fields $L_n$ chosen in such a way that ${\cal M}z$ and $P(z)$
agree order by order in $z$. For the moment we do not specify the domain
over which the agreement occurs. A lower bound on this domain as a
function of $n$ is given in Section \ref{sec:5}.
In parallel with setting the linear 
transformation in $\cal M$ equal to the identity, we have assumed that
to first order in $z$ $\;\;P(z)=z$.

For a vector $v$, regardless of the vector space, we define the
norm $\Vert v\Vert$ by
\begin{equation}
\Vert v\Vert=\max_i|v_i|,
\label{eq:2.6}
\end{equation}
where $|\cdot |$ stands for the modulus. For brevity we denote the norm
of $z$ by $x$, $\;\;x=\Vert z\Vert$. We also define the quantity 
$\alpha_j^{(n)}$ by
\begin{equation}
\alpha_j^{(n)}=\sum_{|r|=n}|a_{rj}^{(n)}|,
\label{eq:2.7}
\end{equation}
and $\alpha_n$ by $\alpha_n=\big\Vert\alpha^{(n)}\big\Vert$. The following 
relation holds:
\begin{equation}
\big\Vert L_n\big\Vert=\max_j{\big |}g_j^{(n)}(z){\big |}=
\max_j{\big |}\sum_{|r|=n}
a_{r j}^{(n)} z^r{\big |}
\le\alpha_n x^n
\label{eq:2.8}
\end{equation}
In the subsequent sections we will obtain a bound for $\alpha_n$ which will
thus enable us to place a bound on $\big\Vert L_n\big\Vert$.

The first step is to write down an expression 
for the coefficients $a_{r j}^{(n)}$
in terms of the coefficients $b_{r j}^{(n)}$. We expand the Lie 
transformations into power series and match terms of the same order in $z$ to
get
\begin{equation}
\sum_{[n]=n-1} {{L_2^{s_2}\ldots L_n^{s_n}}\over{s_2 !\ldots s_n!}} z_k
=\sum_{|r|=n} b_{r k}^{(n)}  z^r.
\label{eq:2.9}
\end{equation}
The symbol $\sum_{[p]=q}$ is defined as a sum over $s_2,\dots,s_p$ with
a condition,
\begin{equation}
\sum_{[p]=q}=\sum_{{s_2,\ldots,s_p\ge 0}\atop{s_2+2 s_3+
\ldots+(p-1) s_p=q}}.
\label{eq:2.10}
\end{equation}
Note that the operators $L_i^{s_i}$ and $L_j^{s_j}\;\;i\ne j$ do not commute,
so the ordering is important.

In Eq. (\ref{eq:2.9}) $s_n$ can take on only the values of 0 and 1. 
Together with
the fact that $L_n z_k=g_k^{(n)}(z)$, this allows us to transform 
Eq. (\ref{eq:2.9}) into a recursion relation for $a_{r k}^{(n)}$,
\begin{mathletters}
\label{eq:2.11}
\begin{equation}
a_{rk}^{(2)}=b_{rk}^{(2)},
\label{eq:2.11a}
\end{equation}
\begin{equation}
a_{rk}^{(n)}=b_{rk}^{(n)}-
{{\partial_z^r}\over{r!}}\sum_{[n-1]=n-1}
{L_2^{s_2}\ldots L_{n-1}^{s_{n-1}}\over s_2!\ldots s_{n-1}!}
z_k;\;\;\;\;n\ge 3,\;\;n\in{\Bbb N},
\label{eq:2.11b}
\end{equation}
\end{mathletters}
where ${{\partial_z^r}\over{r!}}\,{\buildrel\rm def\over =}\,
{\partial_{z_1}^{r_1}\ldots\partial_{z_d}^{r_d}\over
r_1!\ldots r_d!}$.
(Throughout the paper ${\Bbb N}$ is taken to include $0$.)
This is the starting point for the computation of estimates for $\alpha_n$.

\section{A Recursion Relation for Norms}
\label{sec:3}

With the definition
\begin{equation}
\beta_n=\max_j\big(\sum_{|r|=n}|b_{rj}^{(n)}| \big),
\label{eq:3.1}
\end{equation}
Eq. (\ref{eq:2.11b}) yields
\begin{eqnarray}
\alpha_n & = & \max_k \sum_{|r|=n}\Big| b_{rk}^{(n)}-
{{\partial_z^r}\over{r!}}\sum_{[n-1]=n-1}
{L_2^{s_2}\ldots L_{n-1}^{s_{n-1}}\over s_2!\ldots s_{n-1}!}
z_k \Big|\nonumber\\ 
& \le &
\beta_n+\max_k \sum_{|r|=n}\Big|\sum_{[n-1]=n-1}
{{\partial_z^r}\over{r!}}
{L_2^{s_2}\ldots L_{n-1}^{s_{n-1}}\over s_2!\ldots s_{n-1}!} \Big|\nonumber\\
& \le &
\beta_n +\sum_{[n-1]=n-1}{1\over{s_2!\ldots s_{n-1}!}}\max_k
\sum_{|r|=n}\Big| {{\partial_z^r}\over{r!}}
L_2^{s_2}\ldots L_{n-1}^{s_{n-1}} z_k \Big|
\label{eq:3.2}
\end{eqnarray}
Note that the component of the vector on the right side is determined by
the component of $z$ and is labeled here by $k$. Our goal is to express the
right side of the last inequality in (\ref{eq:3.2}) in terms of 
$\alpha_2,\ldots,\alpha_{n-1}$. We accomplish this through two lemmas (the
second one will also be used in Section \ref{sec:5}).

Consider a vector function $F$ whose components are homogeneous polynomials,
\begin{equation}
F_k^{(l)}=\sum_{|i|=l} f_{ik}^{(l)}z^i;\;\;\;\;
1\le k\le d,
\label{eq:3.3}
\end{equation}
and for which
\begin{equation}
\max_k\big( \sum_{|i|=l}| f_{ik}^{(l)}|\big)\le \phi_l
\label{eq:3.4}
\end{equation}
for some $\phi_l\in{\Bbb R}^+$. Define the quantities $c$, $m$, and $B$ by
\begin{mathletters}
\label{eq:3.5}
\begin{equation}
\sum_{|t|=m(\phi_l x^l,s_2,\ldots,s_n)} c_{tk}(F,s_2,\ldots,s_n) 
z^t=
L_2^{s_2}\ldots L_n^{s_n} F_k^{(n)}(z),
\label{eq:3.5a}
\end{equation}
\begin{equation}
B(\phi_l x^l,s_2,\ldots,s_n) x^{m(\phi_l x^l,s_2,\ldots,s_n)}=
\big( x^2\alpha_2 {d\over{dx}}\big)^{s_2}\cdots
\big( x^n\alpha_n {d\over{dx}}\big)^{s_n}\phi_l x^l.
\label{eq:3.5b}
\end{equation}
\end{mathletters}
Here use is made of the obvious fact that the power of $x$ in 
(\ref{eq:3.5b}) is the same as the power of $z$ in (\ref{eq:3.5a}).
Evidently $B(\phi_l x^l,s_2,\ldots,s_n)$ is a nonnegative real quantity and 
$m(\phi_l x^l,s_2,\ldots,s_n)$ is a nonnegative integer. The arguments of
$c$, $m$, and $B$ have been chosen to be rather explicit, so that the 
notation is sufficiently general for the manipulations that follow. 
When referring to a power of only one vector field, on the other hand,
we drop the subscript on the summation index, denoting it by $s$, and 
replace the argument $\phi_l x^l$ of $m$ and $B$ simply by $l$.
The following holds:

{\bf Lemma 3.1.} For all $s\in{\Bbb N}$ and $l\in{\Bbb N}$ 
\begin{equation}
\max_k\big(\sum_{|t|=m(l,s)}|c_{tk}(F,s)| \big)\le B(l,s).
\label{eq:3.6}
\end{equation}
and
\begin{equation}
\big\Vert L_n^s F^{(l)}(z)\big\Vert\le B(l,s) x^{m(l,s)}
\label{eq:3.7}
\end{equation}
This lemma is given in Refs. \cite{grobner}, though its proof is only
outlined there. In the Appendix we provide a more complete 
proof (which is an extension of the relations derived in Appendix A of 
Ref. \cite{me}).

The product of operators of the form $L_i^{s_i}$ can now be bounded by the 
lemma below.

{\bf Lemma 3.2.} For all $n\ge 2$, $n\in{\Bbb N}$ and
all functions $F$ of the form (\ref{eq:3.3}), $l\in{\Bbb N}$,
\begin{equation}
\max_k\sum_{|t|=m(\phi_l x^l,s_2,\ldots,s_n)}|c_{tk}(F,s_2,\ldots,s_n)|\le 
B(\phi_l x^l,s_2,\ldots,s_n)
\label{eq:3.8}
\end{equation}
and
\begin{equation}
\big\Vert L_2^{s_2}\ldots L_n^{s_n} F^{(l)}(z)\big\Vert\le 
B(\phi_l x^l,s_2,\ldots,s_n) x^{m(\phi_l x^l,s_2,\ldots,s_n)}.
\label{eq:3.9}
\end{equation}

Proof is by induction on $n$ and is straightforward. For $n=2$ 
inequalities (\ref{eq:3.8},\ref{eq:3.9}) are the same as inequalities 
(\ref{eq:3.6},\ref{eq:3.7}) and thus hold by Lemma 3.1. Assume now
(\ref{eq:3.8},\ref{eq:3.9}) hold for a fixed $n$ and all functions $F$ of 
the form (\ref{eq:3.3}). Then
\begin{equation}
\big\Vert L_2^{s_2}\ldots L_n^{s_n} L_{n+1}^{s_{n+1}}F^{(l)}(z)\big\Vert =
\big\Vert L_2^{s_2}\ldots L_n^{s_n} \tilde F^{(l')}(z)\big\Vert
\label{eq:3.10}
\end{equation}
where
\begin{equation}
\tilde F_k^{(l')}(z)=\sum_{|t|=l'} c_{tk}(F,s_{n+1}) z^t
\label{eq:3.11}
\end{equation}
with $l'=n s_{n+1}+l$ and, by Lemma 3.1,
\begin{equation}
\max_k\sum_{|t|=l'}|c_{tk}(F,s_{n+1})|\le B(\phi_l x^l,s_{n+1}).
\label{eq:3.12}
\end{equation}
Use of the induction assumption yields
\begin{equation}
\max_k\sum_{|t|=m(B(\phi_l x^l,s_{n+1})x^{l'},s_2,\ldots,s_n)}
|c_{tk}(\tilde F,s_2,\ldots,s_n)|\le 
B(B(\phi_l x^l,s_{n+1})x^{l'},s_2,\ldots,s_n)
\label{eq:3.13}
\end{equation}
and
\begin{equation}
\big\Vert L_2^{s_2}\ldots L_n^{s_n} \tilde F^{(l')}(z)\big\Vert\le
B(B(\phi_l x^l,s_{n+1})x^{l'},s_2,\ldots,s_n) 
x^{m(B(\phi_l x^l,s_{n+1})x^{l'},s_2,\ldots,s_n)}.
\label{eq:3.14}
\end{equation}
By unfolding the definitions of $m$ and $B$ we get
\begin{eqnarray}
&& 
B(B(\phi_l x^l,s_{n+1})x^{l'},s_2,\ldots,s_n) 
x^{m(B(\phi_l x^l,s_{n+1})x^{l'},s_2,\ldots,s_n)} \nonumber\\
&&=
\big( x^2\alpha_2 {d\over{dx}}\big)^{s_2}\cdots
\big( x^n\alpha_n {d\over{dx}}\big)^{s_n} B(\phi_l x^l,s_{n+1}) x^{l'}
\nonumber\\
&&=
\big( x^2\alpha_2 {d\over{dx}}\big)^{s_2}\cdots
\big( x^n\alpha_n {d\over{dx}}\big)^{s_n} 
\big( x^{n+1}\alpha_{n+1} {d\over{dx}}\big)^{s_{n+1}}\phi_l x^l\nonumber\\
&&=
B(\phi_l x^l,s2,\ldots,s_n,s_{n+1}) x^{m(\phi_l x^l,s2,\ldots,s_n,s_{n+1})},
\label{eq:3.15}
\end{eqnarray}
and so replace $B(B(\phi_l x^l,s_{n+1})x^{l'},s_2,\ldots,s_n)$ by 
$B(\phi_l x^l,s_2,\ldots,s_n,s_{n+1})$ and 
$m(B(\phi_l x^l,s_{n+1})x^{l'},s_2,\ldots,s_n)$ by 
$m(\phi_l x^l,s_2,\ldots,s_n,s_{n+1})$ in inequalities 
(\ref{eq:3.13},\ref{eq:3.14}). With the further replacement of 
$c_{tk}({\tilde F},s_2,\ldots,s_n)$ by $c_{tk}(F,s_2,\ldots,s_n,s_{n+1})$,
the proof is complete.
\hfill{$\square$}

We can now make progress with inequality (\ref{eq:3.2}). Since
\begin{equation}
\max_k\sum_{|r|=n}\Big| {{\partial_z^r}\over{r!}}
L_2^{s_2}\ldots L_{n-1}^{s_{n-1}} z_k \Big|=
\max_k\sum_{|r|=n}|c_{rk}(z,s_2,\ldots,s_{n-1})|,
\label{eq:3.16}
\end{equation}
use of inequality (\ref{eq:3.8}) yields
\begin{equation}
\alpha_n\le\beta_n+\sum_{[n-1]=n-1}{1\over{s_2!\ldots s_{n-1}!}}
B(x,s_2,\ldots,s_{n-1}).
\label{eq:3.17}
\end{equation}
Note that the apparent dependence of the right side of (\ref{eq:3.2}) on
$z$ (or $\Vert z\Vert$) has disappeared, as it should.

The final step is to obtain an explicit expression for 
$B(x,s_2,\ldots,s_{n-1})$, which requires the evaluation of the right side
of Eq. (\ref{eq:3.5b}) for $l=1, \;\phi_l=1$. First we note that
\begin{equation}
\Big(x^n {d\over{d x}}\Big)^s x^p={{(n-1)^s\Gamma\Big({{s(n-1)+p}\over{n-1}}
\Big)}\over{\Gamma\Big({p\over{n-1}}\Big)}}x^{p+s(n-1)},
\label{eq:3.18}
\end{equation}
where we take, as it is sufficient for our purposes, $n,s,$ and $p$ to be
integers, with $n\ge2,\;\;s\ge 0,$ and $p\ge 1$. The relation (\ref{eq:3.18})
is easily proven by induction on $s$. Repeated use of (\ref{eq:3.18}) on
the right side of Eq. (\ref{eq:3.5b}) then leads to
\begin{eqnarray}
& &B(x,s_2,\ldots,s_{n-1})=\alpha_2^{s_2} \big(2\alpha_3\big)^{s_3}\cdots
\big( (n-2)\alpha_{n-1}\big)^{s_{n-1}}\nonumber\\
& \times &{{\Gamma\bigg({{1+s_{n-1}(n-2)}
\over{n-2}}\bigg)}\over{\Gamma\bigg({1\over{n-2}}\bigg)}}
{{\Gamma\bigg({{1+s_{n-1}(n-2)+s_{n-2}(n-3)}
\over{n-3}}\bigg)}\over{\Gamma\bigg({{1+s_{n-1}(n-2)}\over{n-3}}\bigg)}}
\cdots
{{\Gamma\bigg({{1+s_{n-1}(n-2)+\cdots+s_2}
\over{1}}\bigg)}\over{\Gamma\bigg({{1+s_{n-1}(n-2)+\cdots+2 s_3}\over{1}}
\bigg)}},
\label{eq:3.19}
\end{eqnarray}
which is the desired, though admittedly cumbersome, expression for $B$.

With the definitions $\eta_n=n \alpha_{n+1}$,
$\;\;\tau_n=n \beta_{n+1}$, 
\begin{mathletters}
\begin{eqnarray}
& &Q_m=1;\hspace{2.07truein} m=n-1\nonumber\\
& &Q_m=1+
\sum_{j=1}^{n-m-1}(n-j) s_{n-j+1};\;\;\;\;1\le m\le n-2,
\label{eq:3.20a}
\end{eqnarray}
and
\begin{equation}
{\cal G}(s_2,\ldots,s_n)=
\prod_{m=1}^{n-1}{{\Gamma\Big(s_{m+1}+{Q_m\over m}\Big)}
\over{s_{m+1}!\;\Gamma\Big({Q_m\over m}\Big)}},
\label{eq:3.20b}
\end{equation}
\end{mathletters}
inequality (\ref{eq:3.17}) and Eq. (\ref{eq:2.11a}) become
\begin{mathletters}
\label{eq:3.21}
\begin{equation}
\eta_1=\tau_1
\label{eq:3.21a}
\end{equation}
\begin{equation}
\eta_n\le\tau_n+n \sum_{[n]=n}\eta_1^{s_2}\ldots\eta_{n-1}^{s_n}
{\cal G}(s_2,\ldots,s_n);\;\;\;\;n\ge 2,\;\;n\in{\Bbb N}.
\label{eq:3.21b}
\end{equation}
\end{mathletters}
In the next section we will use these relations to get a bound for $\eta_n$.
We call attention to the interesting fact that the relations (\ref{eq:3.21}),
and hence the results that follow, do not depend explicitly on $d$. The 
dimensionality of the space enters only through the definition of quantities
$\eta_n$ and $\tau_n$.

\section{Bound for $\eta_n$}
\label{sec:4}

Inequality (\ref{eq:3.21b}) is a complicated relation between $\eta$'s
and $\tau$'s. The reader is invited to show that attempts to establish
simple estimates for $\eta_n$, such as  $\eta_n\le K^n$ or 
$\eta_n\le K^n n!,$ $\;\;K\in {\Bbb R}^+$, by induction from (\ref{eq:3.21b})
fail. (For the latter case note that the sums over $s$
always contain the term  $s_2=s_n=1$, $s_i=0;\;\; 3\le i\le n-1$, for which
${\cal G}(1,0,\ldots,0,1)={n\over {n-1}}$.) The following gives a bound for 
$\eta_n$.

{\bf Theorem 4.1.} Let $K=\max_n \tau_n^{1\over n}$ and define the
quantities $h_n$ by 
\begin{equation}
h_1=1;\;\;\;\; h_n=\prod_{j=2}^n (1+2^{j\over{j-1}}(j-1))^{1\over j}; 
\;\; n\ge 2,\; n\in{\Bbb N}.
\label{eq:4.1}
\end{equation}
Then
\begin{equation}
\eta_n\le K^n h_n^n.
\label{eq:4.2}
\end{equation}

{\sc proof}. First, since $P(z)$ of Eq. (\ref{eq:2.5}) is a convergent power
series, $|b_{rk}^{(n)}|$ is bounded by an exponentially growing function of 
$n$. As
\begin{equation}
\beta_n\le{{n+d-1}\choose{n}}\max_{r k}|b_{rk}^{(n)}|,
\label{eq:4.3}
\end{equation}
$\beta_n$ is also bounded by an exponentially growing function of $n$, and
so is $\tau_n$. Thus the quantity $K=\max_n \tau_n^{1\over n}$ for 
$n\ge 2,\;$ $n\in{\Bbb N}$, exists and is finite.

Next, we manipulate the ratios of $\Gamma$ functions that appear in
$\cal G$. For $1\le m\le n-1$,
\begin{eqnarray}
{{\Gamma(s_{m+1}+{Q_m\over m})}\over{\Gamma({Q_m\over m})}}&=&
\big(s_{m+1}-1+{Q_m\over m}\big)\big(s_{m+1}-2+{Q_m\over m}\big)
\cdots {Q_m\over m}\nonumber\\
&\le &\big(s_{m+1}-1+{Q_m\over m}\big)^{s_{m+1}}\nonumber\\
&=&{1\over{m^{s_{m+1}}}}[m(s_{m+1}-1)+Q_m]^{s_{m+1}}\nonumber\\
&\le&{1\over{m^{s_{m+1}}}}(n+1-m)^{s_{m+1}}.
\label{eq:4.4}
\end{eqnarray}
The last inequality makes use of $m s_{m+1}+Q_m\le n+1$, which follows 
from the condition on the sums over $s$. Hence,
\begin{equation}
{\cal G}(s_2,\ldots,s_n)\le{{n^{s_2} (n-1)^{s_3}\cdots 2^{s_n}}
\over{s_2!\cdots s_n! 1^{s_2}2^{s_3}\cdots (n-1)^{s_n}}}
\label{eq:4.5}
\end{equation}
and inequality (\ref{eq:3.21b}) becomes
\begin{equation}
\eta_n\le\tau_n+n\sum_{[n]=n}{{(n \eta_1)^{s_2}[(n-1)\eta_2]^{s_3}
\cdots(2\eta_{n-1})^{s_n}}\over
{s_2!\cdots s_n! 1^{s_2}\cdots (n-1)^{s_n}}}.
\label{eq:4.6}
\end{equation}
To proceed further we establish the following statement.

{\bf Lemma 4.1.} For $n\ge 2$ and $1\le m\le n-1$
\begin{equation}
(n-m+1)^{1\over m} h_m\le 2^{1\over{n-1}} h_{n-1}.
\label{eq:4.7}
\end{equation}

Proof of Lemma 4.1 is effected in five steps.

(i) For the case $m=n-1$ (\ref{eq:4.7}) obviously holds. Since
$m=n-1$ is the only value of $m$ when $n=2$, it remains to consider
$n\ge 3,\;\;$ $1\le m\le n-2$.

For further manipulations it is useful to denote the ratio of the left and 
right sides of (\ref{eq:4.7}) by $G$,
\begin{equation}
G(n,m)={{(n-m+1)^{1\over m} h_m}
\over{2^{1\over{n-1}} h_{n-1}}}=
{{(n-m+1)^{1\over m}}
\over{2^{1\over{n-1}}
\prod_{j=m+1}^{n-1} (1+2^{j\over{j-1}}(j-1))^{1\over j}}}.
\label{eq:4.8}
\end{equation}
We thus need to show that $G(n,m)\le 1$.

(ii) Let $m=n-2$. For $n=3$ the direct calculation shows that $G(3,1)=0.95$,
whereas for $n\ge 4$ we have
\begin{eqnarray}
G(n,n-2)&=&{{3^{1\over{n-2}}}
\over{2^{1\over{n-1}}[1+2^{{n-1}\over{n-2}}(n-2)]^{1\over{n-1}}}}
< {{3^{1\over{n-2}}}\over{2^{1\over{n-1}}
(1+2^{{n-1}\over{n-2}})^{1\over{n-1}}}}\nonumber\\
&<&\bigg({3\over{2^{{2n-3}\over{n-1}}}}\bigg)^{1\over{n-2}}
\le \Big({3\over{2^{5\over 3}}}\Big)^{1\over{n-2}}=0.94^{1\over{n-2}}.
\label{eq:4.9}
\end{eqnarray}
The last inequality uses ${{2n-3}\over{n-1}}\ge{5\over 3}$ which evidently 
holds for $n\ge 4$. It remains to consider $1\le m\le n-3;$ $\;\;n\ge 4$.

(iii) Let $m=1$. By direct calculation $G(4,1)=0.75$, whereas for $n\ge 5$ we 
manipulate the product appearing in the definition of $G$ to get
\begin{eqnarray}
\prod_{j=2}^{n-1}(1+2^{j\over{j-1}}(j-1))^{1\over j}&=&
\exp[\sum_{j=2}^{n-1}{1\over j}\log(1+2^{j\over{j-1}}(j-1))]
>\exp[\sum_{j=2}^{n-1}{1\over j}\log(2 j-1)]\nonumber\\
&\ge&
\exp[{{n-2}\over{n-1}}\log(2n-3)]=(2n-3)^{{n-2}\over{n-1}}.
\label{eq:4.10}
\end{eqnarray}
The last inequality follows from the monotonic decrease of the summand as
a function of $j$. We provide here a brief justification of this argument
which, with slight modifications, also appears in step (v).
Treating $j$ as a continuous variable, we have
\begin{equation}
{{d\Big[{1\over j}\log(2j-1)\Big]}\over{dj}}=
{{(2j-1)(1-\log(2j-1))+1}\over{j^2(2j-1)}}.
\label{eq:4.11}
\end{equation}
For $j=2$ the value of the numerator is $-0.70$, and this value obviously
decreases with increasing $j$.

Returning to the bounding of $G$, since $n\ge 5$ we write
$2n-3\ge{7\over 5}n$ and use inequality (\ref{eq:4.10}) to get
\begin{equation}
G(n,1)<
\bigg[{1\over{\big({7\over 5}\big)^{n-2} {2\over n}}}\bigg]^{1\over{n-1}}.
\label{eq:4.12}
\end{equation}
Note that if $n$ is treated as a continuous variable then
\begin{equation}
{{d\Big[\big({7\over 5}\big)^{n-2}{2\over n}\Big]}
\over{dn}}={{50\big({7\over 5}\big)^n}\over{49 n}}
\Big[\log\big({7\over 5}\big)-{1\over n}\Big],
\label{eq:4.13}
\end{equation}
and so for $n\ge{1\over{\log\big({7\over 5}\big)}}=2.97$, the denominator is 
a monotonically increasing function of $n$. At $n=5$ it takes the value 
$1.10$. Therefore, $G(n,1)<1$ for $n\ge 5$. It remains to consider
$2\le m\le n-3;$ $\;\;n\ge 5$.

(iv) By direct calculation $G(5,2)=0.52$, and we are left to
explore only $2\le m\le n-3;$ $\;\;n\ge 6$.

(v) Now we can consider the remaining values of $m$ and $n$. Proceeding in 
analogy with Eq. (\ref{eq:4.10}), we write
\begin{equation}
\prod_{j=m+1}^{n-1}(1+2^{j\over{j-1}}(j-1))^{1\over j}
>(2n-3)^{{n-m-1}\over{n-1}}>
n^{{n-m-1}\over{n-1}}.
\label{eq:4.14}
\end{equation}
Therefore,
\begin{equation}
G(n,m)<{{(n-m+1)^{1\over m}}\over{2^{1\over{n-1}}n^{{n-m-1}\over{n-1}}}}
<{1\over{2^{1\over{n-1}}}}n^{{1\over{m(n-1)}}(n-1-m(n-1)+m^2)}.
\label{eq:4.15}
\end{equation}

We now examine the parabola
\begin{equation}
{\cal P}(m)=m^2-m(n-1)+n-1,
\label{eq:4.16}
\end{equation}
where $m$ takes on all real values. Zeroes of ${\cal P}(m)$ are
located at
\begin{equation}
m_{+/-}={1\over 2}\big[n-1\pm\sqrt{(n-1)^2-4 n+4}\big].
\label{eq:4.17}
\end{equation}
The quantity under the square root satisfies
\begin{equation}
n^2-6n+5=(n-4)^2+2n-11>(n-4)^2\;\;\;\;{\rm for}\;\; n\ge6,
\label{eq:4.18}
\end{equation}
and so $m_-<{3\over 2}$ and $m_+>n-{5\over 2}$. Therefore for 
$2\le m\le n-3$, $\;\;{\cal P}(m)<0$, which gives $G(n,m)<1$.

Putting together the results of (i)--(v) completes the proof of Lemma 4.1.
\hfill{$\square$}

To finish the proof of Theorem 4.1 we carry out an induction on $n$. For 
$n=1$ we have
\begin{equation}
\eta_1=\tau_1\le K,
\label{eq:4.19}
\end{equation}
which verifies (\ref{eq:4.2}). Assume now that (\ref{eq:4.2}) holds 
through $n-1$. Then inequality (\ref{eq:4.6}) gives
\begin{eqnarray}
\eta_n&\le&\tau_n+n K^n \sum_{[n]=n} n^{s_2} [(n-1)^{1\over 2} h_2]^{2 s_3}
\cdots [2^{1\over{n-1}}h_{n-1}]^{(n-1)s_n}
{1\over{s_2!\cdots s_n! 1^{s_2}\cdots(n-1)^{s_n}}}\nonumber\\
&\le&
\tau_n+n K^n
\sum_{[n]=n}[2^{1\over{n-1}}h_{n-1}]^{s_2+2s_3+\cdots+(n-1)s_n}
{1\over{s_2!\cdots s_n! 1^{s_2}\cdots(n-1)^{s_n}}}\nonumber\\
&\le&
\tau_n+n K^n
2^{n\over{n-1}}h_{n-1}^n\sum_{[n]=n}
{1\over{s_2!\cdots s_n! 1^{s_2}\cdots(n-1)^{s_n}}},
\label{eq:4.20}
\end{eqnarray}
where we have used the condition on the sum to sum the power of the summand.
The second inequality follows from Lemma 4.1. 
The remaining sums over $s$ nicely sum to $1-{1\over n}$, as can
be demonstrated using Cauchy's identity \cite{riordan}
\begin{equation}
\sum_{[n+1]=n}
{1\over{s_2!\cdots s_n! s_{n+1}!
1^{s_2}\cdots(n-1)^{s_n}n^{s_{n+1}}}}=1.
\label{eq:4.21}
\end{equation}
Substitution of this result into Eq. (\ref{eq:4.20}) leads to the
inequalities
\begin{eqnarray}
\eta_n&\le&\tau_n+K^n 2^{n\over{n-1}} h_{n-1}^n (n-1)
\le K^n\Big[1+2^{n\over{n-1}} h_{n-1}^n (n-1)\Big]\nonumber\\
&\le&
K^n \Big[h_{n-1}\big(1+2^{n\over{n-1}}(n-1)\big)^{1\over n}\Big]^n
=K^n h_n^n.
\label{eq:4.22}
\end{eqnarray}
For the second inequality we have used $K^n\ge\tau_n$ and for the
third one $h_{n-1}\ge 1$. This completes the proof of Theorem 4.1.
\hfill{$\square$}

The large--$n$ behavior of the estimate (\ref{eq:4.2}) is not easy to discern. 
We thus provide an asymptotic expression for the
result. First we convert the product appearing in (\ref{eq:4.1}) into a sum
by taking the logarithm of $h_n^n$, and then use the Euler--Maclaurin Sum
Formula \cite{bender} to get 
\begin{mathletters}
\begin{eqnarray}
\log h_n&=&\sum_{k=0}^m \theta(k)\nonumber\\
&\sim& {1\over 2} \theta(m)+
\int_0^m \theta(t) dt+c_1+\sum_{j=1}^\infty (-1)^{j+1} {B_{j+1}\over{(j+1)!}}
{{d^j \theta(m)}\over{d m}};\;\;\;\;m\rightarrow\infty,
\label{eq:4.23a}
\end{eqnarray}
where
\begin{equation}
\theta(k)={1\over{k+2}}\log\big(1+2^{{k+2}\over{k+1}}(k+1)\big),
\label{eq:4.23b}
\end{equation}
\end{mathletters}
$B_n$ are the Bernoulli numbers, $m=n-2$ and $c_1$ is a constant to be 
determined. The integral of $\theta$ evaluates to
\begin{equation}
\int_0^m \theta(t) dt={1\over 2}\big[\log(m+2)\big]^2+\log 2\log(m+2)
+c_2+{1\over{m+2}}({1\over 2}-\log 2)+O\big({1\over {m^2}}\big),
\label{eq:4.24}
\end{equation}
where $c_2$ is given by $c_2=-0.577765\ldots$. The terms in the sum over $j$,
on the other hand, are of order ${{\log(n)}\over {n^2}}$ and can be 
neglected. Nevertheless, for the numerical computation of the constant
$c_1$ we have used the $j=1$ term in order to improve the numerical
convergence of the procedure (the $j=2$ term is zero since $B_3=0$). The
resulting asymptotic expansion for the bound on $\eta_n$ reads
\begin{equation}
K^n h_n^n\sim \big( K c_3\big)^n \sqrt{{e n}\over 2}
n^{n(\log 2+{1\over 2}\log n)}\times\exp\big[O\big({{\log n}\over n}\big)\big];
\;\;\;\;n\rightarrow\infty.
\label{eq:4.25}
\end{equation}
The constant $c_3$ is given by $c_3=\exp(c_1+c_2)$ and takes the value 
$c_3=0.857695\ldots$.

\section{Optimum Truncation}
\label{sec:5}

In Ref. \cite{me} we have shown that sufficient conditions for convergence
of ${\cal M}_n z$ as $n\rightarrow\infty$ are that $\eta_n$ be bounded by
an exponentially growing function of $n$ and that $z$ be restricted to
a suitable domain around the origin. 
Yet, inequality (\ref{eq:3.21b}) is not consistent with
$\eta_n\le K^n$, $\;\; K\in{\Bbb R}^+$. Instead of considering convergence,
then, we turn to the asymptotic properties of ${\cal M}_n z$. In particular, 
we examine the question of optimum truncation.

The most natural quantity to optimize is the difference between $P(z)$ and
${\cal M}_n z$. Denoting this quantity by $R(n)$, we have
\begin{eqnarray}
R(n)&=&\big\Vert P(z)-e^{L_2}\cdots e^{L_n} z\big\Vert\nonumber\\
&=&
\big\Vert \sum_{q=n+1}^\infty \sum_{|r|=q} b_{r k}^{(q)}
z^r - \sum_{q=n+1}^\infty \sum_{[n]=q-1}
{{L_2^{s_2}\cdots L_n^{s_n}}\over{s_2!\cdots s_n!}} z_k\big\Vert\nonumber\\
&\le&\sum_{q=n+1}^\infty x^q \Big[ \beta_q+\sum_{[n]=q-1} 
{{B(x,s_2,\ldots,s_n)}
\over{s_2!\cdots s_n!}}\Big]\nonumber\\
&=&
\sum_{q=n+1}^\infty x^q\Big[ \beta_q+\sum_{[n]=q-1}
\eta_1^{s_2}\cdots\eta_{n-1}^{s_n} {\cal G}(s_2,\ldots,s_n)\Big].
\label{eq:5.1}
\end{eqnarray}
For the inequality we have used Lemma 3.2 and the condition on the sums 
over s, and for the last equality Eq. (\ref{eq:3.19}). Note that $R(n)$ is
not necessarily defined on the entire domain ${\cal D}$. This question will
be addressed later in this section. We could now substitute
the result of Theorem 4.1 in the last expression in Eq. (\ref{eq:5.1}) to
get an explicit estimate for $R(n)$. The resulting expression, however, 
involves an infinite sum (where each term is very complicated). We thus
use additional inequalities to obtain a closed--form estimate for $R(n)$.

The first to be simplified is the ratio of $\Gamma$ functions occurring in
${\cal G}$. (It is not fruitful to use Eq. (\ref{eq:4.4}) here 
because the condition on the sums over $s$ is different.) For
$1\le m\le n-1$ the following holds:
\begin{eqnarray}
{{\Gamma(s_{m+1}+{{Q_m}\over m})}
\over{\Gamma({{Q_m}\over m})}}
&=&(s_{m+1} -1+{{Q_m}\over m})(s_{m+1}-2+{{Q_m}\over m})
\cdots{{Q_m}\over m}\nonumber\\
&=&{1\over{m^{s_{m+1}}}}\big(m s_{m+1}-m+Q_m\big)\big(m s_{m+1}-2 m+Q_m\big)
\cdots Q_m\nonumber\\
&\le&
{1\over{m^{s_{m+1}}}}(q-m)(q-2 m)\cdots(q-s_{m+1} m)\nonumber\\
&\le&
{1\over{m^{s_{m+1}}}}(q-1)(q-2)\cdots(q-s_{m+1})\nonumber\\
&=&{1\over{m^{s_{m+1}}}}{{(q-1)!}\over{(q-1-s_{m+1})!}}.
\label{eq:5.2}
\end{eqnarray}
The first inequality relies on the relation $m s_{m+1}+Q_m\le q$.
Using (\ref{eq:5.2}) and (\ref{eq:4.2}), the sums over $s$ in 
Eq. (\ref{eq:5.1}) become
\begin{eqnarray}
& &\sum_{[n]=q-1}\eta_1^{s_2}\cdots\eta_{n-1}^{s_n}
{\cal G}(s_2,\ldots,s_n)\nonumber\\
&\le &
\big(K 2^{1\over{n-1}} h_{n-1}\big)^{q-1}
\sum_{[n]=q-1}{{{\cal G}(s_2,\ldots,s_n)}
\over{n^{s_2}(n-1)^{s_3}\cdots 2^{s_n}}}\nonumber\\
&\le&
\big(K 2^{1\over{n-1}} h_{n-1}\big)^{q-1}
\sum_{[n]=q-1}
{{q-1}\choose{s_2}}{1\over{n^{s_2}}}
{{q-1}\choose{s_3}}{1\over{[2(n-1)]^{s_3}}}
\cdots{{q-1}\choose{s_n}}{1\over{[2(n-1)]^{s_n}}}
\nonumber\\
&\le&
\big(K 2^{1\over{n-1}} h_{n-1}\big)^{q-1}
\sum_{s_2=0}^{q-1}\cdots\sum_{s_n=0}^{q-1}
{{q-1}\choose{s_2}}{1\over{n^{s_2}}}
{{q-1}\choose{s_3}}{1\over{[2(n-1)]^{s_3}}}
\cdots{{q-1}\choose{s_n}}{1\over{[2(n-1)]^{s_n}}}
\nonumber\\
&\le&
\big(K 2^{1\over{n-1}} h_{n-1}\big)^{q-1}
\bigg[\prod_{j=1}^{n-1}\Big(1+{1\over{j(n-j+1)}}\Big)\bigg]^{q-1}
\nonumber\\
&\le&
\big(K 2^{1\over{n-1}} h_{n-1}\big)^{q-1}
(1+{1\over n})^{(n-1)(q-1)}\nonumber\\
&<&
\big(K e 2^{1\over{n-1}} h_{n-1}\big)^{q-1}.
\label{eq:5.3}
\end{eqnarray}
The first inequality follows from Lemma 4.1 and the condition on the sum,
the second one from Eq. (\ref{eq:5.2}), the third one is evident upon an
examination of the ranges of indices $s_2$ through $s_n$ subject to the
condition $[n]=q-1$, whereas the last one is clear from the relation
$\big(1+{1\over n}\big)^{n-1}<\big(1+{1\over n}\big)^n<e$. Finally, 
as mentioned in the proof of Theorem 4.1,
$\beta_q$ is bounded by an exponentially growing function of $q$, 
\begin{equation}
\beta_q\le \Big({c\over{3^{1\over 3}}}\Big)^{q-1},
\label{eq:5.4}
\end{equation}
for some $c\in{\Bbb R}^+$. Consistent with Eq. (\ref{eq:2.5}) we have taken 
$\beta_1=1$. In terms of $c$, $\;\;K$ satisfies $K\le c$. 

{}From inequality (\ref{eq:5.3}), and using Eq. (\ref{eq:5.4}), we then obtain 
the following closed--form estimate for $R(n)$:
\begin{mathletters}
\label{eq:5.5}
\begin{eqnarray}
R(n)&<&{{3^{1\over 3}}\over c}\sum_{q=n+1}^\infty 
\Big({{c x}\over{3^{1\over 3}}}\Big)^q+
{1\over{c e 2^{1\over {n-1}} h_{n-1} }}\sum_{q=n+1}^\infty
(c x e 2^{1\over{n-1}} h_{n-1})^q\label{eq:5.5a}\\
&=&{{x^{n+1}c^n}\over{3^{n\over 3}\Big(1-{{c x}\over{3^{1\over 3}}}\Big)}}+
{{x^{n+1}(c e 2^{1\over{n-1}} h_{n-1})^n}\over
{1-c x e 2^{1\over{n-1}} h_{n-1}}}
\,{\buildrel\rm def\over =}\,R^\ast (n).
\label{eq:5.5b}
\end{eqnarray}
\end{mathletters}
Note that Eq. (\ref{eq:5.3}) or Eq. (\ref{eq:5.5a}) explicitly demonstrates
the expected result that regardless of the dependence of $\eta_n$ on $n$,
one can always choose $x$ sufficiently small to guarantee that ${\cal M}_n z$ 
is a convergent transformation (that is that the sums of the form 
(\ref{eq:2.4}) acting on $z$ converge). The condition for convergence is
\begin{equation}
x<{1\over{c e 2^{1\over{n-1}} h_{n-1}}},
\label{eq:5.6}
\end{equation}
which shows the shrinking of the lower bound on the domain of analyticity
of ${\cal M}_n z$ with increasing $n$. (Evidently, (\ref{eq:5.6}) is more
restrictive than $z\in{\cal D}$. For the latter case it is sufficient that
$x<3^{1\over 3}/c$.)

Since the leading order term in ${\cal M} z$ is $z$, which is of order $x$,
it is useful to divide $R^\ast$ by $x$, so that the resulting quantity
$R^\ast/x$ can be compared to one. Note also that $R^\ast/x$
depends on $c$ and $x$ only through the product $c x$, which we denote by 
$\bar x$.

The question of optimum truncation can now be formulated as follows:
given $\bar x$, find the value of $n$ where $R^\ast/x$
reaches its minimum and find the value of the minimum. 
Eq. (\ref{eq:5.5b}), however, is too complicated to carry out the required
calculations analytically. Instead, we have to rely on (straightforward) 
numerical computations. Figure 1 shows the value of $n$ 
where $R^\ast/x$ reaches its minimum as a function of $-\log_{10}({\bar x})$.
This value of $n$ is denoted by $n_{\rm min}$. Figure 2
gives the base-10 logarithm of the value of the minimum, again vs.
$-\log_{10}({\bar x})$.

If we leave rigor aside, we can obtain analytical expressions that closely
approximate the solid curves in Figures 1 and 2.
First we assume that the minimum of $R^\ast/x$ is determined primarily
by terms of lowest order in ${\bar x}$, ${\bar x}^n$ and then neglect 
the term $({\bar x} 3^{-{1\over 3}})^n$ compared with 
$({\bar x} e 2^{1\over{n-1}} h_{n-1})^n$. The latter step is justified when
$(e 2^{1\over{n-1}} h_{n-1})^n\gg 3^{-{n\over 3}}$. We are thus led to 
examine the quantity
\begin{equation}
{r^\ast(n)\over x}\,{\buildrel\rm def\over =}\,
({\bar x} e 2^{1\over{n-1}} h_{n-1})^n.
\label{eq:5.7}
\end{equation}
The location and value of the minimum of $r^\ast/x$ have been determined
numerically and found to be in excellent agreement with the location
and value of the minimum of $R^\ast/x$, the agreement improving with
decreasing ${\bar x}$. To obtain an analytical expression for the minimum of 
$r^\ast/x$, however, additional approximations are needed.

We take the logarithm of ${{r^\ast}\over x}$, consider $n$ a continuous
variable, differentiate with respect to it, and find the location of the 
minimum by setting the result equal to zero. The derivative is
\begin{equation}
{{d\log({{r^\ast}\over x})}\over{d n}}=1+\log{\bar x}
-{{\log 2}\over{(n-1)^2}}+\log h_{n-1}+
n{d\over{d n}}\log h_{n-1}.
\label{eq:5.8}
\end{equation}
In order to evaluate the last term in (\ref{eq:5.8}) we use the identity
\cite{bender}
\begin{equation}
\sum_{k=0}^{n-3} \theta(k)={1\over 2}\big[\theta(0)+\theta(n-3)\big]+
\int_0^{n-3} \theta(t) dt+\int_0^{n-3}(t-[t]-{1\over 2})f'(t) dt,
\label{eq:5.9}
\end{equation}
where $\theta(k)$ is given by Eq. (\ref{eq:4.23b}) and $[t]$ stands for the integer 
part of $t$, which is valid for integer $n$. Then we define $\log h_{n-1}$ 
for 
noninteger $n$ to be the right side of this expression evaluated at noninteger
$n$. Substituting (\ref{eq:5.9}) into (\ref{eq:5.8}), using the asymptotic 
expansion of the form (\ref{eq:4.25}) for the non-differentiated 
$\log h_{n-1}$, and replacing $n-[n]$ by its average value of ${1\over 2}$,
yields
\begin{equation}
{{d\log({{r^\ast}\over x})}\over{d n}}\sim {1\over 2}(\log n)^2+
(1+\log 2)\log n +\log{\bar x}+c_4-{1\over{2 n}}+O\big({{\log n}\over
{n^2}}\big).
\label{eq:5.10}
\end{equation}
The constant $c_4$ is defined by $c_4=\log c_3 +1 +\log 2$. Both this one
and the asymptotic expressions that follow are valid for large values of
$n$. (As is evident from Eq. (\ref{eq:5.11}), and is to be expected, for the 
minimum of ${{r^\ast}\over x}$ this is equivalent to 
${\bar x}\rightarrow 0$.) For brevity we omit writing down explicitly 
$n\rightarrow\infty$ after each asymptotic relation. 

Setting the right side of (\ref{eq:5.10}) equal to zero and neglecting terms of
order ${{\log n}\over{n^2}}$ gives easily
\begin{eqnarray}
n'&\sim& {\rm Int}\Big\{ \exp\big[-1-\log 2+\sqrt{(1+\log 2)^2-2\log{\bar x}
-2 c_4}\big]\nonumber\\
&+&{1\over{2 \sqrt{(1+\log 2)^2-2\log{\bar x}-2 c_4}}}\Big\}.
\label{eq:5.11}
\end{eqnarray}
Here Int stands for the integer closest to the real number enclosed in the 
braces and we have denoted the integer nearest to the zero of 
(\ref{eq:5.10}) by $n'$. 
The right side of Eq. (\ref{eq:5.11}) agrees very well 
with the numerical results obtained for ${{R^\ast}\over x}$: of the 91 
points included in Figure 1, the two functions differ by 1 at only one point.
The exponential term alone of Eq. (\ref{eq:5.11}) also agrees very well with 
the numerical results. We have included the correction, however, to ensure 
that the expression for the value of the minimum is correct through constant 
terms.

The asymptotic expansion for ${{r^\ast}\over x}$ follows from the asymptotic
expansion for $\log h_{n-1}$. The result is
\begin{equation}
{{r^\ast}\over x}\sim ({\bar x} e c_3)^n\sqrt{{e\over{2 n}}}
n^{n(\log 2+{1\over 2}\log n)}\times\exp
\big[O\big({{\log n}\over n}\big)\big].
\label{eq:5.12}
\end{equation}
It is now straightforward to subsitute the right side of Eq. (\ref{eq:5.11})
into the right side of Eq. (\ref{eq:5.12}) and obtain an explicit expression for
the minimum of ${{r^\ast}\over x}$. The result is a lengthy formula which
we do not reproduce here. It provides, however, a good approximation to the 
numerical result. Figure 3 shows the difference between the base-10 logarithm
of this analytical formula and the numerical result, as a function of 
$-\log_{10}{\bar x}$.

Eq. (\ref{eq:5.12}) is valid for any value of $n$, not only at the minimum
of ${{r^\ast}\over x}$. Should we wish to use Eq. (\ref{eq:5.10})
to simplify (\ref{eq:5.12}) at the minimum we could, but care should be
taken to include the fact that $n'$ is the zero of the right side of 
(\ref{eq:5.10}) rounded to the nearest integer. The difference between $n'$ and
the actual zero is $O({1\over n})$, which in ${{r^\ast}\over x}$ gives 
corrections of order one. Here we then merely note that at the minimum
\begin{equation}
{{r^\ast}\over x}\bigg{\vert}_{n=n'}\propto {1\over{\sqrt{n'}
2^{n'} n'^{n'}}}\times\exp\big[O\big({{\log n'}\over n'}\big)\big].
\label{eq:5.13}
\end{equation}

\section{Summary}
\label{sec:6}

In Theorem 4 we have given an upper bound on the norm of vector
fields $L_n$ which are computed by requiring that ${\cal M}z$
agrees order by order with a given convergent power series. The bound
grows with order more rapidly than the
exponential function. Thus we cannot use the results of Ref.\cite{me} to
ascertain the existence of a finite domain in $x$ for which ${\cal M}_n
z$ is convergent and analytic as $n\rightarrow\infty$ 
(analyticity follows from the analytic nesting of domains of
successive Lie transformations -- see Eq. (3.1) in \cite{me}).
Instead, we have sought to optimize the difference between ${\cal M}_n z$
and $P(z)$ as a function of $n$. While an exact analytical
expression for the minimum of a bound on the difference proved elusive, 
at least without significantly weakening the bound or the results of 
Theorem 4, we have given an
asymptotic expression which is valid when the minimum occurs after
a large number of Lie transformations. Comparison of asymptotic 
and numerical results, however, showed close agreement
between the two even for values of ${\bar x}$ for which it is
optimal to use a relatively small number of Lie transformations. (For example,
the results agree to better than 5\% for $n_{\rm min}=6$.)

It seems well worthwhile to explore now whether the procedure developed 
in the preceding sections can be adapted to Hamiltonian normal form 
calculations and used to strengthen the estimates 
of the type given in Ref. \cite{koseleff} for the norm of 
generating polynomials.
It would also be interesting to examine if the absence of
convergence of ${\cal M}_n z$ as $n\rightarrow\infty$ 
is only apparent, due to estimates that were used to
obtain a rigorous bound, or is the true property of Lie
transformations computed from a power series. The first step in this
direction may be the numerical computation of ${\cal M}_n$ through
a large value of $n$ for some representative $P(z)$. Note that the
calculation of coefficients $a_{rk}^{(n)}$, using 
Eq. (\ref{eq:2.11}) or an equivalent, requires only algebraic
manipulations, as all derivatives act on powers of $z$.

\acknowledgments

This work was supported in part by the U.S. Department of Energy
Grant No. DE--FG05--92ER40748.

\appendix

\section*{Proof of Lemma 3.1}

We first prove the relation (\ref{eq:3.6}) by induction. For $s=0$,
$m(l,0)=l$, $c_{tk}(F,0)=f_{tk}^{(l)}$, and $B(l,0)=\phi_l$.
Therefore (\ref{eq:3.6}) reduces to inequality (\ref{eq:3.4}). For
$s=1$ we have
\begin{eqnarray}
L_n F_k^{(l)}(z)&=&\sum_{j=1}^d g_j^{(n)} {\partial\over{\partial z_j}}
\sum_{|i|=l}f_{ik}^{(l)} z^i\nonumber\\
&=&\sum_{j=1}^d\sum_{|r|=n}\sum_{|i|=l}a_{rj}^{(n)}f_{ik}^{(l)}
i_j z_1^{i_1+r_1}\cdots z_j^{i_j+r_j-1}\cdots z_d^{i_d+r_d},
\label{eq:a1}
\end{eqnarray}
which yields
\begin{eqnarray}
\max_k\bigg(\sum_{|t|=m(l,1)}\big|c_{tk}(F,1)\big|\bigg)&\le&
\max_k\bigg(
\sum_{j=1}^d\sum_{|r|=n}\sum_{|i|=l}\big|a_{rj}^{(n)}\big| \big|
f_{ik}^{(l)}\big| i_j\bigg)\nonumber\\
&\le&\alpha_n\max_k\bigg(\sum_{j=1}^d
\sum_{|i|=l}\big|f_{ik}^{(l)}\big| i_j\bigg)\nonumber\\
&=&\alpha_n l 
\max_k\bigg(\sum_{|i|=l}\big|f_{ik}^{(l)}\big|\bigg)\nonumber\\
&\le&\alpha_n l\phi_l\nonumber\\
&=&B(l,1),
\label{eq:a2}
\end{eqnarray}
as needed. Next, assume that (\ref{eq:3.6}) holds for a fixed $s$. Then
\begin{eqnarray}
L_n^{s+1} F_k^{(l)}(z)&=&\sum_{j=1}^d g_j^{(n)} {\partial\over{\partial z_j}}
\sum_{|t|=m(l,s)}c_{tk}(F,s) z^t\nonumber\\
&=&\sum_{j=1}^d\sum_{|r|=n}\sum_{|t|=m(l,s)}a_{rj}^{(n)}c_{tk}(F,s) t_j
z_1^{t_1+r_1}\cdots z_j^{t_j+r_j-1}\cdots z_d^{t_d+r_d},
\label{eq:a3}
\end{eqnarray}
and we have
\begin{eqnarray}
\max_k\bigg(\sum_{|t|=m(l,s+1)}\big|c_{tk}(F,s+1)\big|\bigg)&\le&\max_k\bigg(
\sum_{j=1}^d\sum_{|r|=n}\sum_{|t|=m(l,s)}\big|a_{rj}^{(n)}\big| \big|
c_{tk}(F,s)\big| t_j\bigg)\nonumber\\
&\le &\alpha_n\max_k\bigg(\sum_{j=1}^d
\sum_{|t|=m(l,s)}\big|c_{tk}(F,s)\big| t_j\bigg)\nonumber\\
&=&\alpha_n m(l,s)\max_k\bigg(\sum_{|t|=m(l,s)}\big|c_{tk}(F,s)\big|
\bigg)\nonumber\\
&\le&\alpha_n m(l,s) B(l,s)\nonumber\\
&=&B(l,s+1).
\label{eq:a4}
\end{eqnarray}
The last inequality uses the induction assumption, whereas the last equality
follows from the recursion relation satisfied by $B$. This completes the 
proof of relation (\ref{eq:3.6}).

It is now straightforward to establish (\ref{eq:3.7}). We proceed 
again by induction. For $s=0$ we use inequality (\ref{eq:3.4}) and the special
values of $m, c,$ and $B$ given at the beginning of the Appendix to see that
(\ref{eq:3.7}) holds. For the case $s=1$, on the other hand, we use relations
(\ref{eq:a1}) and (\ref{eq:a2}) (second inequality)
to get 
\begin{eqnarray}
\Vert L_n F_k^{(l)}(z)\Vert&=&\Vert\sum_{j=1}^d\sum_{|r|=n}
\sum_{|i|=l}a_{rj}^{(n)}f_{ik}^{(l)}
i_j z_1^{i_1+r_1}\cdots z_j^{i_j+r_j-1}\cdots z_d^{i_d+r_d}\Vert\nonumber\\
&\le&\max_k\bigg(
\sum_{j=1}^d\sum_{|r|=n}\sum_{|i|=l}\big|a_{rj}^{(n)}\big| \big|
f_{ik}^{(l)}\big| i_j\bigg)x^{n+l-1}\nonumber\\
&\le& B(l,1) x^{m(l,1)}.
\label{eq:a5}
\end{eqnarray}
The first inequality is evident from the definition of the norm and for the 
last relation we have relied on the fact that $m(l,1)=n+l-1$. Assume now
that (\ref{eq:3.7}) holds for a fixed $s$. With the help of relations
(\ref{eq:a3}) and (\ref{eq:a4}) (second inequality) we get
\begin{eqnarray}
\Vert L_n^{s+1} F_k^{(l)}(z)\Vert&=&\Vert\sum_{j=1}^d\sum_{|r|=n}
\sum_{|t|=m(l,s)}a_{rj}^{(n)}c_{tk}(F,s) t_j
z_1^{t_1+r_1}\cdots z_j^{t_j+r_j-1}\cdots z_d^{t_d+r_d}\Vert\nonumber\\
&\le&\max_k\bigg(
\sum_{j=1}^d\sum_{|r|=n}\sum_{|t|=m(l,s)}\big|a_{rj}^{(n)}\big| \big|
c_{tk}(F,s)\big| t_j\bigg)x^{m(l,s)+n-1}\nonumber\\
&\le& B(l,s+1)x^{m(l,s+1)}.
\label{eq:a6}
\end{eqnarray}
We have again made use of the recursion relation for $m$, $m(l,s+1)=
m(l,s)+n-1$. This completes the proof of inequality (\ref{eq:3.7}).


\newpage

\centerline{\bf Figure Captions}

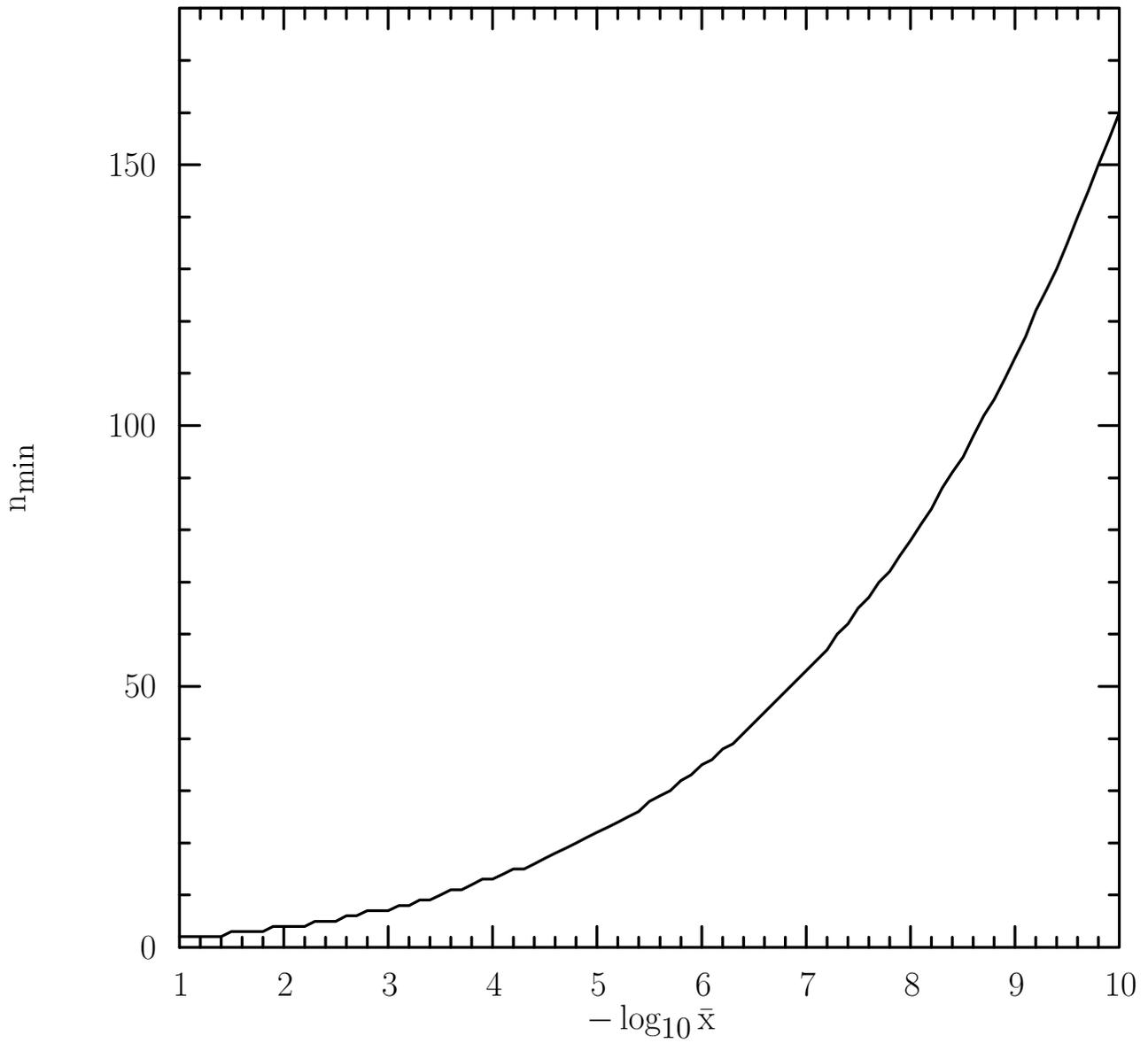
\noindent Figure 1. $n_{\rm min}$ vs. $-\log_{10}{\bar x}$. The step size
in $\log_{10}{\bar x}$ is 0.1.

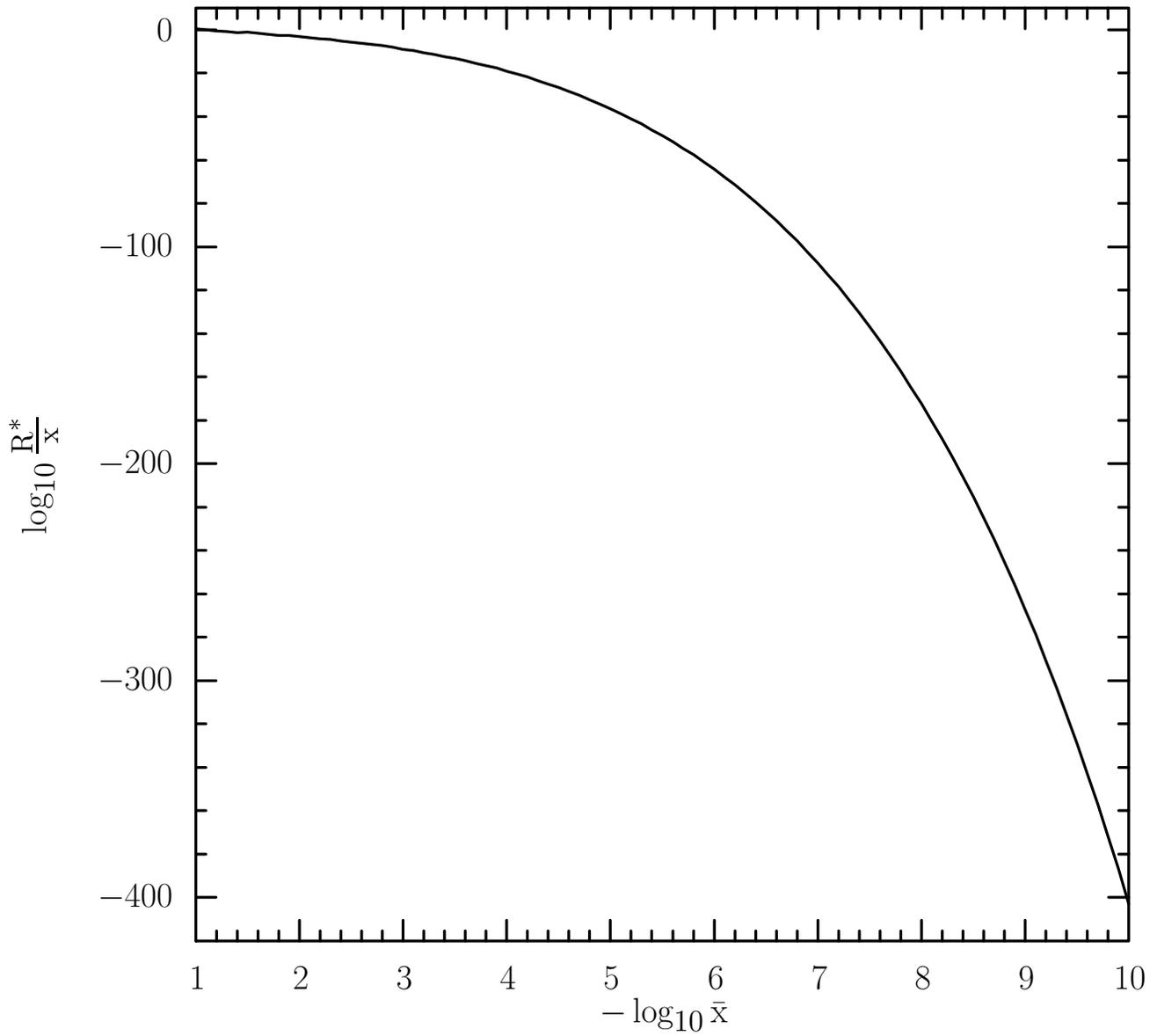
\noindent Figure 2. The value of $\log_{10}R^\ast/x$ at $n=n_{\rm min}$ 
vs. $-\log_{10}{\bar x}$. The step size in $\log_{10}{\bar x}$ is 0.1, as
in Figure 1.

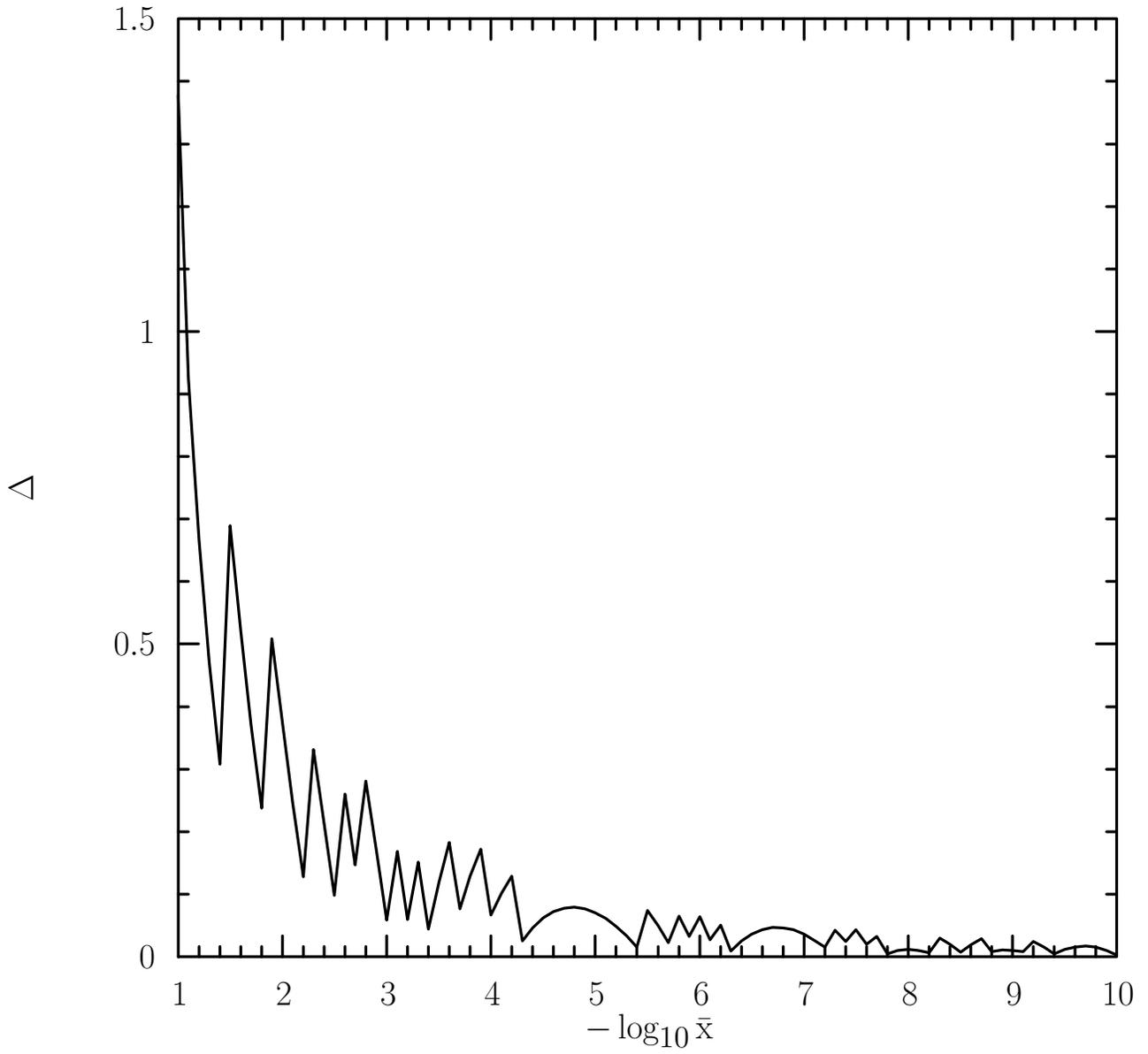
\noindent Figure 3. The quantity $\Delta({\bar x})=
\log_{10}R^\ast/x\Big{\vert}_{n=n_{\rm min}}-
\log_{10}{{r^\ast_a}/x}\Big{\vert}_{n=n_a'}$ vs. $-\log_{10}{\bar x}$. Here 
${{r^\ast_a}/x}$ denotes the right side of Eq. (\ref{eq:5.12}) and 
$n_a'$ the right side of Eq. (\ref{eq:5.11}).
The step size in $\log_{10}{\bar x}$ is 0.1, as in Figures 1 and 2.

\vfill\eject
\ifx\axisloaded\relax \fi






 
\def\setRevDate $#1 #2 #3${\def\TeXdrawId{TeXdraw V1R3 revised <#2>}}
\setRevDate $Date: Wed Mar  3 12:01:12 MST 1993$

\chardef\catamp=\the\catcode`\@
\catcode`\@=11
\ifx\TeXdraw@included\undefined\global\let\TeXdraw@included=\relax\else
\errhelp{TeXdraw needs to be input only once outside of any groups.}%
\errmessage{Multiple call to include TeXdraw ignored}%
\expandafter \fi

\long                              
\def\centertexdraw #1{\hbox to \hsize{\hss
                                      \btexdraw #1\etexdraw
                                      \hss}}


\def\btexdraw {\x@pix=0             \y@pix=0
               \x@segoffpix=\x@pix  \y@segoffpix=\y@pix
               \t@exdrawdef
               \setbox\t@xdbox=\vbox\bgroup\offinterlineskip
                   \global\d@bs=0           
                   \t@extonlytrue           
                   \p@osinitfalse
                   \savemove \x@pix \y@pix  
                   \m@pendingfalse
                   \p@osinitfalse           
                   \p@athfalse}


\def\etexdraw {\ift@extonly \else
                 \t@drclose      
               \fi
               \egroup           
               \ifdim \wd\t@xdbox>0pt
                 \errmessage{TeXdraw box non-zero size,
                             possible extraneous text}%
               \fi
               \maxhvpos         
               \pixtodim \xminpix \l@lxpos  \pixtodim \yminpix \l@lypos
               \pixtobp {-\xminpix}\l@lxbp  \pixtobp {-\yminpix}\l@lybp
               \vbox {
		      \offinterlineskip
                      \ift@extonly \else
                        \includepsfile{\p@sfile}{\the\l@lxbp}{\the\l@lybp}%
				      {\the\hdrawsize}{\the\vdrawsize}%
                      \fi
                      \vskip\vdrawsize
                      \vskip \l@lypos
                      \hbox {\hskip -\l@lxpos
                             \box\t@xdbox
                             \hskip \hdrawsize
                             \hskip \l@lxpos}%
                      \vskip -\l@lypos\relax}}

%

\def\drawdim #1 {\def\d@dim{#1\relax}}


\def\setunitscale #1 {\edef\u@nitsc{#1}%
                      \realmult \u@nitsc  \s@egsc \d@sc}
\def\relunitscale #1 {\realmult {#1}\u@nitsc \u@nitsc
                      \realmult \u@nitsc \s@egsc \d@sc}
\def\setsegscale #1 {\edef\s@egsc {#1}%
                     \realmult \u@nitsc \s@egsc \d@sc}
\def\relsegscale #1 {\realmult {#1}\s@egsc \s@egsc
                     \realmult \u@nitsc \s@egsc \d@sc}

\def\bsegment {\ifp@ath
                 \flushmove
               \fi
               \begingroup
               \x@segoffpix=\x@pix
               \y@segoffpix=\y@pix
               \setsegscale 1
               \global\advance \d@bs by 1 }
\def\esegment {\endgroup
               \ifnum \d@bs=0
                 \writetx {es}%
               \else
                 \global\advance \d@bs by -1
               \fi}

\def\savecurrpos (#1 #2){\getsympos (#1 #2)\a@rgx\a@rgy
                         \s@etcsn \a@rgx {\the\x@pix}%
                         \s@etcsn \a@rgy {\the\y@pix}}%
\def\savepos (#1 #2)(#3 #4){\getpos (#1 #2)\a@rgx\a@rgy
                            \coordtopix \a@rgx \t@pixa
                            \advance \t@pixa by \x@segoffpix
                            \coordtopix \a@rgy \t@pixb
                            \advance \t@pixb by \y@segoffpix
                            \getsympos (#3 #4)\a@rgx\a@rgy
                            \s@etcsn \a@rgx {\the\t@pixa}%
                            \s@etcsn \a@rgy {\the\t@pixb}}

\def\linewd #1 {\coordtopix {#1}\t@pixa
                \flushbs
                \writetx {\the\t@pixa\space sl}}
\def\setgray #1 {\flushbs
                 \writetx {#1 sg}}
\def\lpatt (#1){\listtopix (#1)\p@ixlist
                \flushbs
                \writetx {[\p@ixlist] sd}}

\def\lvec (#1 #2){\getpos (#1 #2)\a@rgx\a@rgy
                  \s@etpospix \a@rgx \a@rgy
                  \writeps {\the\x@pix\space \the\y@pix\space lv}}
\def\rlvec (#1 #2){\getpos (#1 #2)\a@rgx\a@rgy
                   \r@elpospix \a@rgx \a@rgy
                   \writeps {\the\x@pix\space \the\y@pix\space lv}}
\def\move (#1 #2){\getpos (#1 #2)\a@rgx\a@rgy
                  \s@etpospix \a@rgx \a@rgy
                  \savemove \x@pix \y@pix}
\def\rmove (#1 #2){\getpos (#1 #2)\a@rgx\a@rgy
                   \r@elpospix \a@rgx \a@rgy
                   \savemove \x@pix \y@pix}

\def\lcir r:#1 {\coordtopix {#1}\t@pixa
                \writeps {\the\t@pixa\space cr}%
                \r@elupd \t@pixa \t@pixa
                \r@elupd {-\t@pixa}{-\t@pixa}}
\def\fcir f:#1 r:#2 {\coordtopix {#2}\t@pixa
                     \writeps {#1 \the\t@pixa\space fc}%
                     \r@elupd \t@pixa \t@pixa
                     \r@elupd {-\t@pixa}{-\t@pixa}}
\def\lellip rx:#1 ry:#2 {\coordtopix {#1}\t@pixa
                         \coordtopix {#2}\t@pixb
                         \writeps {\the\t@pixa\space \the\t@pixb\space el}%
                         \r@elupd \t@pixa \t@pixb
                         \r@elupd {-\t@pixa}{-\t@pixb}}
\def\larc r:#1 sd:#2 ed:#3 {\coordtopix {#1}\t@pixa
                            \writeps {\the\t@pixa\space #2 #3 ar}}


\def\ifill f:#1 {\writeps {#1 fl}}     
\def\lfill f:#1 {\writeps {#1 fp}}     



\def\htext #1{\def\testit {#1}%
              \ifx \testit\l@paren
                \let\next=\h@move
              \else
                \let\next=\h@text
              \fi
              \next{#1}}

\def\rtext td:#1 #2{\def\testit {#2}%
                    \ifx \testit\l@paren
                      \let\next=\r@move
                    \else
                      \let\next=\r@text
                    \fi
                    \next td:#1 {#2}}

\def\vtext {\rtext td:90 }

\def\textref h:#1 v:#2 {\ifx #1R%
                          \edef\l@stuff {\hss}\edef\r@stuff {}%
                        \else
                          \ifx #1C%
                            \edef\l@stuff {\hss}\edef\r@stuff {\hss}%
                          \else  
                            \edef\l@stuff {}\edef\r@stuff {\hss}%
                          \fi
                        \fi
                        \ifx #2T%
                          \edef\t@stuff {}\edef\b@stuff {\vss}%
                        \else
                          \ifx #2C%
                            \edef\t@stuff {\vss}\edef\b@stuff {\vss}%
                          \else  
                            \edef\t@stuff {\vss}\edef\b@stuff {}%
                          \fi
                        \fi}

\def\avec (#1 #2){\getpos (#1 #2)\a@rgx\a@rgy
                  \s@etpospix \a@rgx \a@rgy
                  \writeps {\the\x@pix\space \the\y@pix\space (\a@type)
                            \the\a@lenpix\space \the\a@widpix\space av}}

\def\ravec (#1 #2){\getpos (#1 #2)\a@rgx\a@rgy
                   \r@elpospix \a@rgx \a@rgy
                   \writeps {\the\x@pix\space \the\y@pix\space (\a@type)
                             \the\a@lenpix\space \the\a@widpix\space av}}

\def\arrowheadsize l:#1 w:#2 {\coordtopix{#1}\a@lenpix
                              \coordtopix{#2}\a@widpix}
\def\arrowheadtype t:#1 {\edef\a@type{#1}}

\def\clvec (#1 #2)(#3 #4)(#5 #6)%
           {\getpos (#1 #2)\a@rgx\a@rgy
            \coordtopix \a@rgx\t@pixa
            \advance \t@pixa by \x@segoffpix
            \coordtopix \a@rgy\t@pixb
            \advance \t@pixb by \y@segoffpix
            \getpos (#3 #4)\a@rgx\a@rgy
            \coordtopix \a@rgx\t@pixc
            \advance \t@pixc by \x@segoffpix
            \coordtopix \a@rgy\t@pixd
            \advance \t@pixd by \y@segoffpix
            \getpos (#5 #6)\a@rgx\a@rgy
            \s@etpospix \a@rgx \a@rgy
            \writeps {\the\t@pixa\space \the\t@pixb\space 
                      \the\t@pixc\space \the\t@pixd\space 
                      \the\x@pix\space \the\y@pix\space cv}}
\def\e@tendspline#1\endpoints{}
\newtoks\splinet@ks
\def\splinep@int #1 #2 %
           {%
            \advance\p@intnumber by 1\relax
	    \splinet@ks={(#1 #2)}%
            \us@rconvert (#1 #2)\a@rgx\a@rgy
            \futurelet\n@xttok\wr@tesplinepoint}
\def\wr@tesplinepoint{%
            \ifx\n@xttok\endpoints
              \s@etpospix \a@rgx \a@rgy
	      \expandafter
	      \us@rfinish \expandafter
              \p@intnumber \expandafter:\the\splinet@ks=({\x@pix} {\y@pix})%
              \expandafter\e@tendspline
            \else
              \coordtopix \a@rgx\t@pixa
              \advance \t@pixa by \x@segoffpix
              \coordtopix \a@rgy\t@pixb
              \advance \t@pixb by \y@segoffpix
	      \expandafter
              \us@rpoint \expandafter
              \p@intnumber \expandafter:\the\splinet@ks=({\t@pixa} {\t@pixb})%
              \expandafter\splinep@int
            \fi}
\def\defaultus@rfinish#1:(#2 #3)=(#4 #5){\writeps {\the#4 \the#5 \the#1 BSpl}}
\def\defaultus@rpoint#1:(#2 #3)=(#4 #5){\writeps {\the#4 \the#5}}
\def\s@tuseroptions#1/#2/#3/#4@{%
\let\us@rconvert=#1\relax\ifx\us@rconvert\relax\let\us@rconvert=\getpos\fi
\let\us@rpoint=#2\relax\ifx\us@rpoint\relax\let\us@rpoint=\defaultus@rpoint\fi
\let\us@rfinish=#3\relax\ifx\us@rfinish\relax\let\us@rfinish=\defaultus@rfinish
   \fi}%
\def\dooverpoints#1\points{%
    \p@intnumber=0\relax\s@tuseroptions#1///@\splinep@int}
\def\spline{\dooverpoints\points}

\newcount\p@intnumber
\def\curvytype#1{\def\curv@type{#1}}\curvytype{4}%
\def\curvyheight#1{\def\curv@height{#1}}\curvyheight{10}%
\def\curvylength#1{\def\curv@length{#1}}\curvylength{10}%
\def\drawcurvyphoton around (#1 #2) from (#3 #4) to (#5 #6)%
 {\getpos (#1 #2)\a@rgx\a@rgy
  \coordtopix \a@rgx \t@pixa \advance \t@pixa by \x@segoffpix
  \coordtopix \a@rgy \t@pixb \advance \t@pixb by \y@segoffpix
  \writeps {mark \the\t@pixa\space \the\t@pixb}%
  \getpos (#3 #4)\a@rgx\a@rgy
  \s@etpospix \a@rgx\a@rgy
  \writeps {\the\x@pix\space \the\y@pix}%
  \getpos (#5 #6)\a@rgx\a@rgy
  \s@etpospix \a@rgx\a@rgy
  \writeps {\the\x@pix\space \the\y@pix}%
  \writeps {\curv@height\space \curv@length\space \curv@type\space%
              curvyphoton}%
}%
\def\drawcurvygluon around (#1 #2) from (#3 #4) to (#5 #6)%
 {\getpos (#1 #2)\a@rgx\a@rgy
  \coordtopix \a@rgx \t@pixa \advance \t@pixa by \x@segoffpix
  \coordtopix \a@rgy \t@pixb \advance \t@pixb by \y@segoffpix
  \writeps {mark \the\t@pixa\space \the\t@pixb}%
  \getpos (#3 #4)\a@rgx\a@rgy
  \s@etpospix \a@rgx\a@rgy
  \writeps {\the\x@pix\space \the\y@pix}%
  \getpos (#5 #6)\a@rgx\a@rgy
  \s@etpospix \a@rgx\a@rgy
  \writeps {\the\x@pix\space \the\y@pix}%
  \writeps {\curv@height\space 2 mul \curv@length\space \curv@type\space%
              curvygluon}%
}%
\def\blobfreq#1{\def\bl@bfreq{#1}}\blobfreq{0.2}%
\def\blobangle#1{\def\bl@bangle{#1}}\blobangle{0}%
\def\hatchedblob#1{\def\bl@btype{(#1)}}\hatchedblob{B}%
\def\grayblob#1{\def\bl@btype{#1}}%
\def\drawblob xsize:#1 ysize:#2 at (#3 #4)%
{\getpos (#3 #4)\a@rgx\a@rgy \s@etpospix \a@rgx\a@rgy
 \writeps{\the\x@pix\space \the\y@pix}%
 \getpos (#1 #2)\a@rgx\a@rgy 
 \coordtopix \a@rgx\t@pixa \coordtopix\a@rgy\t@pixb \writeps
{\the\t@pixa\space\the\t@pixb\space\bl@bangle\space\bl@bfreq\space\bl@btype}%
 \writeps {blob}%
 \rmove (-#1 -#2)\rmove (#1 #2)\rmove (#1 #2)\rmove (-#1 -#2)}%
\def\drawbb {\bsegment
               \drawdim bp
               \setunitscale 0.24
               \linewd 1           
               \writeps {\the\xminpix\space \the\yminpix\space mv}%
               \writeps {\the\xminpix\space \the\ymaxpix\space lv}%
               \writeps {\the\xmaxpix\space \the\ymaxpix\space lv}%
               \writeps {\the\xmaxpix\space \the\yminpix\space lv}%
               \writeps {\the\xminpix\space \the\yminpix\space lv}%
             \esegment}


\def\getpos (#1 #2)#3#4{\g@etargxy #1 #2 {} \\#3#4%
                        \c@heckast #3%
                        \ifa@st
                          \g@etsympix #3\t@pixa
                          \advance \t@pixa by -\x@segoffpix
                          \pixtocoord \t@pixa #3
                        \fi
                        \c@heckast #4%
                        \ifa@st
                          \g@etsympix #4\t@pixa
                          \advance \t@pixa by -\y@segoffpix
                          \pixtocoord \t@pixa #4
                        \fi}

\def\getsympos (#1 #2)#3#4{\g@etargxy #1 #2 {} \\#3#4%
                           \c@heckast #3%
                           \ifa@st \else
                             \errmessage {TeXdraw: invalid symbolic coordinate}
                           \fi
                           \c@heckast #4%
                           \ifa@st \else
                             \errmessage {TeXdraw: invalid symbolic coordinate}
                           \fi}

\def\listtopix (#1)#2{\def #2{}%
                      \edef\l@ist {#1 }
                      \t@countc=0
                      \loop
                        \expandafter\g@etitem \l@ist \\\a@rgx\l@ist
                        \a@pppix \a@rgx #2
                        \ifx \l@ist\empty
                          \t@countc=1
                        \fi
                      \ifnum \t@countc=0
                      \repeat}


\def\realmult #1#2#3{\dimen0=#1pt
                     \dimen2=#2\dimen0
                     \edef #3{\expandafter\c@lean\the\dimen2}}

\def\intdiv #1#2#3{\t@counta=#1
                   \t@countb=#2
	           \ifnum \t@countb<0
                      \t@counta=-\t@counta
                      \t@countb=-\t@countb
                   \fi
                   \t@countd=1                    
                   \ifnum \t@counta<0
                      \t@counta=-\t@counta
                      \t@countd=-1
                   \fi
	           \t@countc=\t@counta  \divide \t@countc by \t@countb
                   \t@counte=\t@countc  \multiply \t@counte by \t@countb
                   \advance \t@counta by -\t@counte
	           \t@counte=-1
                   \loop
                     \advance \t@counte by 1
	             \ifnum \t@counte<16
                       \multiply \t@countc by 2           
                       \multiply \t@counta by 2           
                       \ifnum \t@counta<\t@countb \else   
                         \advance \t@countc by 1          
                         \advance \t@counta by -\t@countb 
                       \fi
                   \repeat
	           \divide \t@countb by 2         
	           \ifnum \t@counta<\t@countb     
                     \advance \t@countc by 1
                   \fi
                   \ifnum \t@countd<0             
                     \t@countc=-\t@countc
                   \fi
                   \dimen0=\t@countc sp           
                   \edef #3{\expandafter\c@lean\the\dimen0}}

\outer\def\gnewif #1{\count@\escapechar \escapechar\m@ne
  \expandafter\expandafter\expandafter
   \edef\@if #1{true}{\global\let\noexpand#1=\noexpand\iftrue}%
  \expandafter\expandafter\expandafter
   \edef\@if #1{false}{\global\let\noexpand#1=\noexpand\iffalse}%
  \@if#1{false}\escapechar\count@} 
\def\@if #1#2{\csname\expandafter\if@\string#1#2\endcsname}
{\uccode`1=`i \uccode`2=`f \uppercase{\gdef\if@12{}}} 


\def\coordtopix #1#2{\dimen0=#1\d@dim
                     \dimen2=\d@sc\dimen0
                     \t@counta=\dimen2              
                     \t@countb=\s@ppix
                     \divide \t@countb by 2
                     \ifnum \t@counta<0             
                       \advance \t@counta by -\t@countb
                     \else
                       \advance \t@counta by \t@countb
                     \fi
                     \divide \t@counta by \s@ppix
                     #2=\t@counta}

\def\pixtocoord #1#2{\t@counta=#1%
                     \multiply \t@counta by \s@ppix
                     \dimen0=\d@sc\d@dim
                     \t@countb=\dimen0
                     \intdiv \t@counta \t@countb #2}

\def\pixtodim #1#2{\t@countb=#1%
                   \multiply \t@countb by \s@ppix
                   #2=\t@countb sp\relax}

\def\pixtobp #1#2{\dimen0=\p@sfactor pt
                  \t@counta=\dimen0
                  \multiply \t@counta by #1%
                  \ifnum \t@counta < 0             
                    \advance \t@counta by -32768
                  \else
                    \advance \t@counta by 32768
                  \fi
                  \divide \t@counta by 65536
                  #2=\t@counta}
                  
\newcount\t@counta    \newcount\t@countb   
\newcount\t@countc    \newcount\t@countd
\newcount\t@counte
\newcount\t@pixa      \newcount\t@pixb     
\newcount\t@pixc      \newcount\t@pixd
\let\l@lxbp=\t@pixa   \let\l@lybp=\t@pixb  
\let\u@rxbp=\t@pixc   \let\u@rybp=\t@pixd

\newdimen\t@xpos      \newdimen\t@ypos
\let\l@lxpos=\t@xpos  \let\l@lypos=\t@ypos

\newcount\xminpix      \newcount\xmaxpix
\newcount\yminpix      \newcount\ymaxpix

\newcount\a@lenpix     \newcount\a@widpix

\newcount\x@pix        \newcount\y@pix
\newcount\x@segoffpix  \newcount\y@segoffpix
\newcount\x@savepix    \newcount\y@savepix

\newcount\s@ppix       

\newcount\d@bs

\newcount\t@xdnum
\global\t@xdnum=0

\newdimen\hdrawsize    \newdimen\vdrawsize

\newbox\t@xdbox

\newwrite\drawfile

\newif\ifm@pending
\newif\ifp@ath
\newif\ifa@st
\gnewif \ift@extonly
\gnewif\ifp@osinit

\def\l@paren{(}
\def\a@st{*}

\catcode`\%=12
  \def\p@b {
\catcode`\%=14
\catcode`\{=12  \catcode`\}=12  \catcode`\u=1 \catcode`\v=2
  \def\l@br u{v  \def\r@br u}v
\catcode `\{=1  \catcode`\}=2   \catcode`\u=11 \catcode`\v=11

{\catcode`\p=12 \catcode`\t=12
 \gdef\c@lean #1pt{#1}}

\def\sppix#1/#2 {\dimen0=1#2 \s@ppix=\dimen0
                 \t@counta=#1%
                 \divide \t@counta by 2
                 \advance \s@ppix by \t@counta
                 \divide \s@ppix by #1
                 \t@counta=\s@ppix
                 \multiply \t@counta by 65536       
                 \advance \t@counta by 32891        
                 \divide \t@counta by 65782         
                 \dimen0=\t@counta sp
                 \edef\p@sfactor {\expandafter\c@lean\the\dimen0}}

\def\g@etargxy #1 #2 #3 #4\\#5#6{\def #5{#1}%
                                 \ifx #5\empty
                                   \g@etargxy #2 #3 #4 \\#5#6
                                 \else
                                   \def #6{#2}%
                                   \def\next {#3}%
                                   \ifx \next\empty \else
                                     \errmessage {TeXdraw: invalid coordinate}%
                                   \fi
                                 \fi}

\def\c@heckast #1{\expandafter
                  \c@heckastll #1\\}
\def\c@heckastll #1#2\\{\def\testit {#1}%
                        \ifx \testit\a@st
                          \a@sttrue
                        \else
                          \a@stfalse
                        \fi}

\def\g@etsympix #1#2{\expandafter
                     \ifx \csname #1\endcsname \relax
                       \errmessage {TeXdraw: undefined symbolic coordinate}%
                     \fi
                     #2=\csname #1\endcsname}

\def\s@etcsn #1#2{\expandafter
                  \xdef\csname#1\endcsname {#2}}

\def\g@etitem #1 #2\\#3#4{\edef #4{#2}\edef #3{#1}}
\def\a@pppix #1#2{\edef\next {#1}%
                  \ifx \next\empty \else
                    \coordtopix {#1}\t@pixa
                    \ifx #2\empty
                      \edef #2{\the\t@pixa}%
                    \else
                      \edef #2{#2 \the\t@pixa}%
                    \fi
                  \fi}

\def\s@etpospix #1#2{\coordtopix {#1}\x@pix
                     \advance \x@pix by \x@segoffpix
                     \coordtopix {#2}\y@pix
                     \advance \y@pix by \y@segoffpix
                     \u@pdateminmax \x@pix \y@pix}

\def\r@elpospix #1#2{\coordtopix {#1}\t@pixa
                     \advance \x@pix by \t@pixa
                     \coordtopix {#2}\t@pixa
                     \advance \y@pix by \t@pixa
                     \u@pdateminmax \x@pix \y@pix}

\def\r@elupd #1#2{\t@counta=\x@pix
                  \advance\t@counta by #1%
                  \t@countb=\y@pix
                  \advance\t@countb by #2%
                  \u@pdateminmax \t@counta \t@countb}

\def\u@pdateminmax #1#2{\ifnum #1>\xmaxpix
                          \global\xmaxpix=#1%
                        \fi
                        \ifnum #1<\xminpix
                          \global\xminpix=#1%
                        \fi
                        \ifnum #2>\ymaxpix
                          \global\ymaxpix=#2%
                        \fi
                        \ifnum #2<\yminpix
                          \global\yminpix=#2%
                        \fi}

\def\maxhvpos {\t@pixa=\xmaxpix
               \advance \t@pixa by -\xminpix
               \pixtodim  \t@pixa {\dimen2}%
               \global\hdrawsize=\dimen2
               \t@pixa=\ymaxpix
               \advance \t@pixa by -\yminpix
               \pixtodim \t@pixa {\dimen2}%
               \global\vdrawsize=\dimen2\relax}

\def\savemove #1#2{\x@savepix=#1\y@savepix=#2%
                   \m@pendingtrue
                   \ifp@osinit \else
                     \p@osinittrue
                     \global\xminpix=\x@savepix \global\yminpix=\y@savepix
                     \global\xmaxpix=\x@savepix \global\ymaxpix=\y@savepix
                   \fi}

\def\flushmove {\p@osinittrue
                \ifm@pending
                  \writetx {\the\x@savepix\space \the\y@savepix\space mv}%
                  \m@pendingfalse
                  \p@athfalse
                \fi}

\def\flushbs {\loop
                \ifnum \d@bs>0
                  \writetx {bs}%
                  \global\advance \d@bs by -1
              \repeat}
               
\def\h@move #1#2 #3)#4{\move (#2 #3)%
                       \h@text {#4}}
\def\h@text #1{\pixtodim \x@pix \t@xpos
               \pixtodim \y@pix \t@ypos
               \vbox to 0pt{\normalbaselines
                            \t@stuff
                            \kern -\t@ypos
                            \hbox to 0pt{\l@stuff
                                         \kern \t@xpos
                                         \hbox {#1}%
                                         \kern -\t@xpos
                                         \r@stuff}%
                            \kern \t@ypos
                            \b@stuff\relax}}

\def\r@move td:#1 #2#3 #4)#5{\move (#3 #4)%
                             \r@text td:#1 {#5}}
\def\r@text td:#1 #2{\pixtodim \x@pix \t@xpos
                     \pixtodim \y@pix \t@ypos
                     \vbox to 0pt{\kern -\t@ypos
                                  \hbox to 0pt{\kern \t@xpos
                                               \rottxt{#1}{#2}%
                                               \hss}%
                                  \vss}}

\def\rottxt #1#2{\rotsclTeX{#1}{1}{1}{\z@sb{#2}}}%
\def\z@sb #1{\vbox to 0pt{\normalbaselines
                          \t@stuff
                          \hbox to 0pt{\l@stuff
                                       \hbox {#1}%
                                       \r@stuff}%
                          \b@stuff}}

\def\t@exdrawdef {\sppix 300/in            
                  \drawdim in              
                  \edef\u@nitsc {1}
                  \setsegscale 1           
                  \arrowheadsize l:0.16 w:0.08
                  \arrowheadtype t:T
                  \textref h:L v:B }


\def\writeps #1{\flushbs
                \flushmove
                \p@athtrue
                \writetx {#1}}
\def\writetx #1{\ift@extonly
                  \t@extonlyfalse
                  \t@dropen
                \fi
                \w@rps {#1}}
\def\w@rps #1{\immediate\write\drawfile {#1}}

\def\t@dropen {%
  \global\advance \t@xdnum by 1
  \ifnum \t@xdnum<10
    \xdef\p@sfile {\jobname.ps\the\t@xdnum}
  \else
    \xdef\p@sfile {\jobname.p\the\t@xdnum}
  \fi
  \immediate\openout\drawfile=\p@sfile
  \w@rps {\p@b PS-Adobe-3.0 EPSF-3.0}%
  \w@rps {\p@p BoundingBox: (atend)}%
  \w@rps {\p@p Title: TeXdraw drawing: \p@sfile}%
  \w@rps {\p@p Pages: 1 1}%
  \w@rps {\p@p Creator: TeXdraw V1R3}%
  \w@rps {\p@p CreationDate: \the\year/\the\month/\the\day}%
  \w@rps {\p@p DocumentSuppliedResources: ProcSet TeXDraw 2.2 2}%
  \w@rps {\p@p DocumentData: Clean7Bit}%
  \w@rps {\p@p EndComments}%
  \w@rps {\p@p BeginDefaults}%
  \w@rps {\p@p PageNeededResources: ProcSet TeXDraw 2.2 2}%
  \w@rps {\p@p EndDefaults}%
  \w@rps {\p@p BeginProlog}%
  \w@rps {\p@p BeginResource: ProcSet TeXDraw 2.2 2 14696 10668}%
  \w@rps {\p@p VMlocation: local}%
  \w@rps {\p@p VMusage: 14696 10668}%
  \w@rps { /product where}%
  \w@rps {  {pop product (ghostscript) eq /setglobal {pop} def} if}%
  \w@rps { /setglobal where}%
  \w@rps {  {pop currentglobal false setglobal} if}%
  \w@rps { /setpacking where}%
  \w@rps {  {pop currentpacking false setpacking} if}%
  \w@rps {29 dict dup begin}%
  \w@rps {62 dict dup begin}%
  \w@rps { /rad 0 def /radx 0 def /rady 0 def /svm matrix def}%
  \w@rps { /hhwid 0 def /hlen 0 def /ah 0 def /tipy 0 def}%
  \w@rps { /tipx 0 def /taily 0 def /tailx 0 def /dx 0 def}%
  \w@rps { /dy 0 def /alen 0 def /blen 0 def}%
  \w@rps { /i 0 def /y1 0 def /x1 0 def /y0 0 def /x0 0 def}%
  \w@rps { /movetoNeeded 0 def}%
  \w@rps { /y3 0 def /x3 0 def /y2 0 def /x2 0 def}%
  \w@rps { /p1y 0 def /p1x 0 def /p2y 0 def /p2x 0 def}%
  \w@rps { /p0y 0 def /p0x 0 def /p3y 0 def /p3x 0 def}%
  \w@rps { /n 0 def /y 0 def /x 0 def}%
  \w@rps { /anglefactor 0 def /elemlength 0 def /excursion 0 def}%
  \w@rps { /endy 0 def /endx 0 def /beginy 0 def /beginx 0 def}%
  \w@rps { /centery 0 def /centerx 0 def /startangle 0 def }%
  \w@rps { /startradius 0 def /endradius 0 def /elemcount 0 def}%
  \w@rps { /smallincrement 0 def /angleincrement 0 def /radiusincrement 0 def}%
  \w@rps { /ifleft false def /ifright false def /iffill false def}%
  \w@rps { /freq 1 def /angle 0 def /yrad 0 def /xrad 0 def /y 0 def /x 0 def}%
  \w@rps { /saved 0 def}%
  \w@rps {end}%
  \w@rps {/dbdef {1 index exch 0 put 0 begin bind end def}}%
  \w@rps {  dup 3 4 index put dup 5 4 index put bind def pop}%
  \w@rps {/bdef {bind def} bind def}%
  \w@rps {/mv {stroke moveto} bdef}%
  \w@rps {/lv {lineto} bdef}%
  \w@rps {/st {currentpoint stroke moveto} bdef}%
  \w@rps {/sl {st setlinewidth} bdef}%
  \w@rps {/sd {st 0 setdash} bdef}%
  \w@rps {/sg {st setgray} bdef}%
  \w@rps {/bs {gsave} bdef /es {stroke grestore} bdef}%
  \w@rps {/cv {curveto} bdef}%
  \w@rps {/cr \l@br 0 begin}%
  \w@rps { gsave /rad exch def currentpoint newpath rad 0 360 arc}%
  \w@rps { stroke grestore end\r@br\space 0 dbdef}%
  \w@rps {/fc \l@br 0 begin}%
  \w@rps { gsave /rad exch def setgray currentpoint newpath}%
  \w@rps { rad 0 360 arc fill grestore end\r@br\space 0 dbdef}%
  \w@rps {/ar {gsave currentpoint newpath 5 2 roll arc stroke grestore} bdef}%
  \w@rps {/el \l@br 0 begin gsave /rady exch def /radx exch def}%
  \w@rps { svm currentmatrix currentpoint translate}%
  \w@rps { radx rady scale newpath 0 0 1 0 360 arc}%
  \w@rps { setmatrix stroke grestore end\r@br\space 0 dbdef}%
  \w@rps {/fl \l@br gsave closepath setgray fill grestore}%
  \w@rps { currentpoint newpath moveto\r@br\space bdef}%
  \w@rps {/fp \l@br gsave closepath setgray fill grestore}%
  \w@rps { currentpoint stroke moveto\r@br\space bdef}%
  \w@rps {/av \l@br 0 begin /hhwid exch 2 div def /hlen exch def}%
  \w@rps { /ah exch def /tipy exch def /tipx exch def}%
  \w@rps { currentpoint /taily exch def /tailx exch def}%
  \w@rps { /dx tipx tailx sub def /dy tipy taily sub def}%
  \w@rps { /alen dx dx mul dy dy mul add sqrt def}%
  \w@rps { /blen alen hlen sub def}%
  \w@rps { gsave tailx taily translate dy dx atan rotate}%
  \w@rps { (V) ah ne {blen 0 gt {blen 0 lineto} if} {alen 0 lineto} ifelse}%
  \w@rps { stroke blen hhwid neg moveto alen 0 lineto blen hhwid lineto}%
  \w@rps { (T) ah eq {closepath} if}%
  \w@rps { (W) ah eq {gsave 1 setgray fill grestore closepath} if}%
  \w@rps { (F) ah eq {fill} {stroke} ifelse}%
  \w@rps { grestore tipx tipy moveto end\r@br\space 0 dbdef}%
  \w@rps {/setupcurvy \l@br 0 begin}%
  \w@rps { dup 0 eq {1 add} if /anglefactor exch def}%
  \w@rps { abs dup 0 eq {1 add} if /elemlength exch def /excursion exch def}%
  \w@rps { /endy exch def /endx exch def}%
  \w@rps { /beginy exch def /beginx exch def}%
  \w@rps { /centery exch def /centerx exch def}%
  \w@rps { cleartomark}%
  \w@rps { /startangle beginy centery sub beginx centerx sub atan def}%
  \w@rps { /startradius beginy centery sub dup mul }%
  \w@rps {              beginx centerx sub dup mul add sqrt def}%
  \w@rps { /endradius endy centery sub dup mul }%
  \w@rps {            endx centerx sub dup mul add sqrt def}%
  \w@rps { endradius startradius sub }%
  \w@rps { endy centery sub endx centerx sub atan }%
  \w@rps { startangle 2 copy le {exch 360 add exch} if sub dup}%
  \w@rps { elemlength startradius endradius add atan dup add}%
  \w@rps { div round abs cvi dup 0 eq {1 add} if}%
  \w@rps { dup /elemcount exch def }%
  \w@rps { div dup anglefactor div dup /smallincrement exch def}%
  \w@rps { sub /angleincrement exch def}%
  \w@rps { elemcount div /radiusincrement exch def}%
  \w@rps { gsave newpath}%
  \w@rps { startangle dup cos startradius mul }%
  \w@rps { centerx add exch }%
  \w@rps { sin startradius mul centery add moveto}%
  \w@rps { end \r@br 0 dbdef}%
  \w@rps {/curvyphoton \l@br 0 begin}%
  \w@rps { setupcurvy}%
  \w@rps { elemcount \l@br /startangle startangle smallincrement add def}%
  \w@rps {            /startradius startradius excursion add def}%
  \w@rps {            startangle dup cos startradius mul }%
  \w@rps {            centerx add exch }%
  \w@rps {            sin startradius mul centery add}%
  \w@rps {	      /excursion excursion neg def}%
  \w@rps {	      /startangle startangle angleincrement add }%
  \w@rps {                        smallincrement sub def}%
  \w@rps {	      /startradius startradius radiusincrement add def}%
  \w@rps {	      startangle dup cos startradius mul }%
  \w@rps {	      centerx add exch }%
  \w@rps {            sin startradius mul centery add}%
  \w@rps {	      /startradius startradius excursion add def}%
  \w@rps {            /startangle startangle smallincrement add def}%
  \w@rps {             startangle dup cos startradius mul }%
  \w@rps {	       centerx add exch }%
  \w@rps {             sin startradius mul centery add curveto\r@br repeat}%
  \w@rps {	       stroke grestore end}%
  \w@rps {	      \r@br 0 dbdef}%
  \w@rps {/curvygluon \l@br 0 begin}%
  \w@rps { setupcurvy /radiusincrement radiusincrement 2 div def}%
  \w@rps { elemcount \l@br startangle angleincrement add dup}%
  \w@rps {            cos startradius mul centerx add exch}%
  \w@rps {            sin startradius mul centery add}%
  \w@rps {            /startradius startradius radiusincrement add}%
  \w@rps {                         excursion sub def}%
  \w@rps {            startangle angleincrement add dup}%
  \w@rps {            cos startradius mul centerx add exch}%
  \w@rps {            sin startradius mul centery add}%
\w@rps{            startangle angleincrement smallincrement add 2 div add dup}%
  \w@rps {            cos startradius mul centerx add exch}%
  \w@rps {            sin startradius mul centery add}%
  \w@rps {	    curveto}%
\w@rps{      /startangle startangle angleincrement smallincrement add add def}%
  \w@rps {            startangle angleincrement sub dup}%
  \w@rps {            cos startradius mul centerx add exch}%
  \w@rps {            sin startradius mul centery add}%
  \w@rps {	    /startradius startradius radiusincrement add}%
  \w@rps {			excursion add def}%
  \w@rps {            startangle angleincrement sub dup}%
  \w@rps {            cos startradius mul centerx add exch}%
  \w@rps {            sin startradius mul centery add}%
  \w@rps {            startangle dup}%
  \w@rps {            cos startradius mul centerx add exch}%
  \w@rps {            sin startradius mul centery add}%
  \w@rps {	    curveto\r@br repeat}%
  \w@rps { stroke grestore end}%
  \w@rps { \r@br 0 dbdef}%
  \w@rps {/blob \l@br}%
  \w@rps {0 begin st gsave}%
  \w@rps {dup type dup}%
  \w@rps {/stringtype eq}%
  \w@rps {\l@br pop 0 get }%
  \w@rps {dup (B) 0 get eq dup 2 index}%
  \w@rps {(L) 0 get eq or /ifleft exch def}%
  \w@rps {exch (R) 0 get eq or /ifright exch def}%
  \w@rps {/iffill false def \r@br}%
  \w@rps {\l@br /ifleft false def}%
  \w@rps {/ifright false def}%
  \w@rps {/booleantype eq }%
  \w@rps {{/iffill exch def}}%
  \w@rps {{setgray /iffill true def} ifelse \r@br}%
  \w@rps {ifelse}%
  \w@rps {/freq exch def}%
  \w@rps {/angle exch def}%
  \w@rps {/yrad  exch def}%
  \w@rps {/xrad  exch def}%
  \w@rps {/y exch def}%
  \w@rps {/x exch def}%
  \w@rps {newpath}%
  \w@rps {svm currentmatrix pop}%
  \w@rps {x y translate 	}%
  \w@rps {angle rotate}%
  \w@rps {xrad yrad scale}%
  \w@rps {0 0 1 0 360 arc}%
  \w@rps {gsave 1 setgray fill grestore}%
  \w@rps {gsave svm setmatrix stroke grestore}%
  \w@rps {gsave iffill {fill} if grestore}%
  \w@rps {clip newpath}%
  \w@rps {gsave }%
  \w@rps {ifleft  \l@br -3 freq 3 { -1 moveto 2 2 rlineto} for}%
  \w@rps {svm setmatrix stroke\r@br if }%
  \w@rps {grestore}%
  \w@rps {ifright \l@br 3 freq neg -3 { -1 moveto -2 2 rlineto} for}%
  \w@rps {svm setmatrix stroke\r@br if}%
  \w@rps {grestore end}%
  \w@rps {\r@br 0 dbdef}%
  \w@rps {/BSpl \l@br}%
  \w@rps { 0 begin}%
  \w@rps { storexyn}%
  \w@rps { currentpoint newpath moveto}%
  \w@rps { n 1 gt \l@br}%
  \w@rps {  0 0 0 0 0 0 1 1 true subspline}%
  \w@rps {  n 2 gt \l@br}%
  \w@rps {   0 0 0 0 1 1 2 2 false subspline}%
  \w@rps {   1 1 n 3 sub \l@br}%
  \w@rps {    /i exch def}%
  \w@rps {    i 1 sub dup i dup i 1 add dup i 2 add dup false subspline}%
  \w@rps {    \r@br for}%
  \w@rps {   n 3 sub dup n 2 sub dup n 1 sub dup 2 copy false subspline}%
  \w@rps {   \r@br if}%
  \w@rps {  n 2 sub dup n 1 sub dup 2 copy 2 copy false subspline}%
  \w@rps {  \r@br if}%
  \w@rps { end}%
  \w@rps { \r@br 0 dbdef}%
  \w@rps {/midpoint \l@br}%
  \w@rps { 0 begin}%
  \w@rps { /y1 exch def}%
  \w@rps { /x1 exch def}%
  \w@rps { /y0 exch def}%
  \w@rps { /x0 exch def}%
  \w@rps { x0 x1 add 2 div}%
  \w@rps { y0 y1 add 2 div}%
  \w@rps { end}%
  \w@rps { \r@br 0 dbdef}%
  \w@rps {/thirdpoint \l@br}%
  \w@rps { 0 begin}%
  \w@rps { /y1 exch def}%
  \w@rps { /x1 exch def}%
  \w@rps { /y0 exch def}%
  \w@rps { /x0 exch def}%
  \w@rps { x0 2 mul x1 add 3 div}%
  \w@rps { y0 2 mul y1 add 3 div}%
  \w@rps { end}%
  \w@rps { \r@br 0 dbdef}%
  \w@rps {/subspline \l@br}%
  \w@rps { 0 begin}%
  \w@rps { /movetoNeeded exch def}%
  \w@rps { y exch get /y3 exch def}%
  \w@rps { x exch get /x3 exch def}%
  \w@rps { y exch get /y2 exch def}%
  \w@rps { x exch get /x2 exch def}%
  \w@rps { y exch get /y1 exch def}%
  \w@rps { x exch get /x1 exch def}%
  \w@rps { y exch get /y0 exch def}%
  \w@rps { x exch get /x0 exch def}%
  \w@rps { x1 y1 x2 y2 thirdpoint}%
  \w@rps { /p1y exch def}%
  \w@rps { /p1x exch def}%
  \w@rps { x2 y2 x1 y1 thirdpoint}%
  \w@rps { /p2y exch def}%
  \w@rps { /p2x exch def}%
  \w@rps { x1 y1 x0 y0 thirdpoint}%
  \w@rps { p1x p1y midpoint}%
  \w@rps { /p0y exch def}%
  \w@rps { /p0x exch def}%
  \w@rps { x2 y2 x3 y3 thirdpoint}%
  \w@rps { p2x p2y midpoint}%
  \w@rps { /p3y exch def}%
  \w@rps { /p3x exch def}%
  \w@rps { movetoNeeded \l@br p0x p0y moveto \r@br if}%
  \w@rps { p1x p1y p2x p2y p3x p3y curveto}%
  \w@rps { end}%
  \w@rps { \r@br 0 dbdef}%
  \w@rps {/storexyn \l@br}%
  \w@rps { 0 begin}%
  \w@rps { /n exch def}%
  \w@rps { /y n array def}%
  \w@rps { /x n array def}%
  \w@rps { n 1 sub -1 0 \l@br}%
  \w@rps {  /i exch def}%
  \w@rps {  y i 3 2 roll put}%
  \w@rps {  x i 3 2 roll put}%
  \w@rps {  \r@br for end}%
  \w@rps { \r@br 0 dbdef}%
  \w@rps {/bop \l@br save 0 begin /saved exch def end}%
  \w@rps { scale setlinecap setlinejoin setlinewidth setdash moveto}%
  \w@rps { \r@br 1 dbdef}%
  \w@rps {/eop {stroke 0 /saved get restore showpage} 1 dbdef}%
  \w@rps {end /defineresource where}%
  \w@rps { {pop mark exch /TeXDraw exch /ProcSet defineresource cleartomark}}%
  \w@rps { {/TeXDraw exch readonly def} ifelse}%
  \w@rps {/setpacking where {pop setpacking} if}%
  \w@rps {/setglobal where {pop setglobal} if}%
  \w@rps {\p@p EndResource}%
  \w@rps {\p@p EndProlog}%
  \w@rps {\p@p Page: 1 1}%
  \w@rps {\p@p PageBoundingBox: (atend)}%
  \w@rps {\p@p BeginPageSetup}%
  \w@rps {/TeXDraw /findresource where}%
  \w@rps { {pop /ProcSet findresource}}%
  \w@rps { {load} ifelse}%
  \w@rps {begin}%
  \w@rps {0 0 [] 0 3 1 1 \p@sfactor\space \p@sfactor\space bop}%
  \w@rps {\p@p EndPageSetup}%
}

\def\t@drclose {%
  \pixtobp \xminpix \l@lxbp  \pixtobp \yminpix \l@lybp
  \pixtobp \xmaxpix \u@rxbp  \pixtobp \ymaxpix \u@rybp
  \w@rps {\p@p PageTrailer}%
  \w@rps {\p@p PageBoundingBox: \the\l@lxbp\space \the\l@lybp\space
                            \the\u@rxbp\space \the\u@rybp}%
  \w@rps {eop end}%
  \w@rps {\p@p Trailer}%
  \w@rps {\p@p BoundingBox: \the\l@lxbp\space \the\l@lybp\space
                            \the\u@rxbp\space \the\u@rybp}%
  \w@rps {\p@p EOF}%
  \closeout\drawfile
}

\catcode`\@=\catamp
\def\dvialwsetup{
\def\includepsfile##1##2##3##4##5{\special{Insert ##1\space%
                                 }}%
\def\rotsclTeX##1##2##3##4{\special{Insert /dev/null do %
                              3 index exch translate cleartomark %
                              matrix currentmatrix aload pop %
                              7 6 roll restore matrix astore %
                              matrix currentmatrix exch setmatrix %
                              0 0 moveto setmatrix %
                              gsave currentpoint 2 copy translate ##1 rotate %
                              ##2 ##3 scale neg exch neg exch translate %
                              save}%
                   ##4%
                   \special{Insert /dev/null do cleartomark restore %
                           {currentpoint} stopped grestore {moveto} save}}%
}
\def\dvipssetup{
\def\includepsfile##1##2##3##4##5{\vbox to 0pt{%
                             \vskip##5%
			     \includegraphics{##1}%
                             \vss}}
\def\rotsclTeX##1##2##3##4{%
		       ##4
}
}
\dvipssetup

\expandafter\ifx\csname fonts are loaded\endcsname\relax\else \fi
\immediate\openin0 localfonts.tex
\ifeof0\relax \else\ifx\localfontsloaded\donotdefinethis\else \fi
\font\seventeenrm=cmr17 \font\twelverm=cmr12 \font\tenrm=cmr10 
\font\ninerm=cmr9       \font\eightrm=cmr8   \font\sevenrm=cmr7
\font\sixrm=cmr6        \font\fiverm=cmr5
\font\twentyfourrm=cmr17 at 24pt \font\twentyrm=cmr17 at 20pt
\font\sixteenrm=cmr17 at 16pt \font\fourteenrm=cmr12 at 14pt
\font\twelvei=cmmi12    \font\teni=cmmi10    \font\ninei=cmmi9
\font\eighti=cmmi8      \font\seveni=cmmi7   \font\sixi=cmmi6
\font\fivei=cmmi5
\font\twentyfouri=cmmi12 at 24pt \font\twentyi=cmmi12 at 20pt
\font\sixteeni=cmmi12 at 16pt \font\fourteeni=cmmi12 at 14pt
\font\tensy=cmsy10      \font\ninesy=cmsy9   \font\eightsy=cmsy8
\font\sevensy=cmsy7     \font\sixsy=cmsy6    \font\fivesy=cmsy5
\skewchar\tensy='60 \skewchar\ninesy='60 \skewchar\eightsy='60
\skewchar\sevensy='60 \skewchar\sixsy='60 \skewchar\fivesy='60
\font\tenex=cmex10      \font\nineex=cmex9   \font\eightex=cmex8
\font\sevenex=cmex7
\font\fiveex=cmex7 at 5pt
\font\twelveit=cmti12   \font\tenit=cmti10   \font\nineit=cmti9
\font\eightit=cmti8     \font\sevenit=cmti7
\font\twentyfourit=cmti12 at 24pt \font\twentyit=cmti12 at 20pt
\font\sixteenit=cmti12 at 16pt \font\fourteenit=cmti12 at 14pt
\font\fiveit=cmti7 at 5pt
\font\twelvesl=cmsl12   \font\tensl=cmsl10   \font\ninesl=cmsl9
\font\eightsl=cmsl8
\font\twentyfoursl=cmsl12 at 24pt \font\twentysl=cmsl12 at 20pt
\font\sixteensl=cmsl12 at 16pt \font\fourteensl=cmsl12 at 16pt
\font\sevensl=cmsl8 at 7pt \font\fivesl=cmsl8 at 5pt
\font\twelvett=cmtt12   \font\tentt=cmtt10   \font\ninett=cmtt9
\font\eighttt=cmtt8
\hyphenchar\twelvett=-1 \hyphenchar\tentt=-1 \hyphenchar\ninett=-1
\hyphenchar\eighttt=-1
\font\twentyfourtt=cmtt12 at 24pt \font\twentytt=cmtt12 at 20pt
\font\sixteentt=cmtt12 at 16pt \font\fourteentt=cmtt12 at 16pt
\font\seventt=cmtt8 at 7pt \font\fivett=cmtt8 at 5pt
\hyphenchar\twentyfourtt=-1 \hyphenchar\twentytt=-1
\hyphenchar\sixteentt=-1 \hyphenchar\fourteentt=-1
\hyphenchar\seventt=-1 \hyphenchar\fivett=-1
\font\tenbf=cmb10
\font\seventeenss=cmss17\font\twelvess=cmss12\font\tenss=cmss10
\font\niness=cmss9      \font\eightss=cmss8
\font\twentyfourss=cmss17 at 24pt \font\twentyss=cmss17 at 20pt
\font\sixteenss=cmss17 at 16pt \font\fourteenss=cmss12 at 14pt
\font\sevenss=cmss8 at 7pt \font\fivess=cmss8 at 5pt
\font\tenmib=cmmib10    \font\ninemib=cmmib9 \font\eightmib=cmmib8
\font\sevenmib=cmmib7   \font\sixmib=cmmib6  \font\fivemib=cmmib5
\skewchar\tenmib='177   \skewchar\ninemib='177 \skewchar\eightmib='177
\skewchar\sevenmib='177 \skewchar\sixmib='177  \skewchar\fivemib='177
\font\tenbsy=cmbsy10    \font\ninebsy=cmbsy9 \font\eightbsy=cmbsy8
\font\sevenbsy=cmbsy7   \font\sixbsy=cmbsy6  \font\fivebsy=cmbsy5
\skewchar\tenbsy='60 \skewchar\ninebsy='60 \skewchar\eightbsy='60
\skewchar\sevenbsy='60 \skewchar\sixbsy='60 \skewchar\fivebsy='60
\let\localfontsloaded=\relax
\fi
\newif\ifplain
\plainfalse
\def\loadfonts#1#2{%
 \expandafter\ifx\csname#1rm\endcsname\relax
 \expandafter\global\expandafter\font\csname#1rm\endcsname=cmr10 at#2pt \fi
 \expandafter\ifx\csname#1i\endcsname\relax
 \expandafter\global\expandafter\font\csname#1i\endcsname =cmmi10 at#2pt 
 \skewchar\csname#1i\endcsname ='177 \fi
 \expandafter\ifx\csname#1sy\endcsname\relax
 \expandafter\global\expandafter\font\csname#1sy\endcsname=cmsy10 at#2pt 
 \skewchar\csname#1sy\endcsname= '60 \fi
 \expandafter\ifx\csname#1ex\endcsname\relax
 \expandafter\global\expandafter\font\csname#1ex\endcsname=cmex10 at#2pt \fi
 \expandafter\ifx\csname#1it\endcsname\relax
 \expandafter\global\expandafter\font\csname#1it\endcsname=cmti10 at#2pt \fi
 \expandafter\ifx\csname#1sl\endcsname\relax
 \expandafter\global\expandafter\font\csname#1sl\endcsname=cmsl10 at#2pt \fi
 \expandafter\ifx\csname#1tt\endcsname\relax
 \expandafter\global\expandafter\font\csname#1tt\endcsname=cmtt10 at#2pt 
 \hyphenchar\csname#1tt\endcsname=  -1 \fi
 \expandafter\ifx\csname#1bf\endcsname\relax
 \expandafter\global\expandafter\font\csname#1bf\endcsname=cmb10  at#2pt \fi
 \expandafter\ifx\csname#1ss\endcsname\relax
 \expandafter\global\expandafter\font\csname#1ss\endcsname=cmss10 at#2pt \fi
 \expandafter\ifx\csname#1mib\endcsname\relax
 \expandafter\global\expandafter\font\csname#1mib\endcsname=cmmib10 at #2pt 
 \skewchar\csname#1mib\endcsname='177 \fi
 \expandafter\ifx\csname#1bsy\endcsname\relax
 \expandafter\global\expandafter\font\csname#1bsy\endcsname=cmbsy10 at #2pt 
 \skewchar\csname#1bsy\endcsname= '60 \fi
}

\ifplain 
\expandafter\ifx\csname tenmib\endcsname\relax\global\font\tenmib=cmmib10 \fi
\expandafter\ifx\csname ninemib\endcsname\global\font\ninemib=cmmib10 at9pt\fi
\expandafter\ifx\csname sevenmi\endcsname\global\font\sevenmib=cmmib10 at7pt\fi
\expandafter\ifx\csname fivemib\endcsname\global\font\fivemib=cmmib10 at5pt\fi
\expandafter\ifx\csname tenbsy\endcsname\global\font\tenbsy=cmbsy10 \fi
\expandafter\ifx\csname ninebsy\endcsname\global\font\ninebsy=cmbsy10 at9pt\fi
\expandafter\ifx\csname sevenbs\endcsname\global\font\sevenbsy=cmbsy10 at7pt\fi
\expandafter\ifx\csname fivebsy\endcsname\global\font\fivebsy=cmbsy10 at5pt\fi
\else
\def\tenfonts{%
	\loadfonts{ten}{10}%
	\global\def\tenfonts{}}
\def\ninefonts{%
	\loadfonts{nine}{9}%
	\global\def\ninefonts{}}
\def\sevenfonts{%
	\loadfonts{seven}{7}%
	\global\def\sevenfonts{}}
\def\fivefonts{%
	\loadfonts{five}{5}%
	\global\def\fivefonts{}}
\fi
\def\twelvefonts{%
	\loadfonts{twelve}{12}%
	\global\def\twelvefonts{}}
\def\fourteenfonts{%
	\loadfonts{fourteen}{14}%
	\global\def\fourteenfonts{}}
\def\sixteenfonts{%
	\loadfonts{sixteen}{16}%
	\global\def\sixteenfonts{}}
\def\twentyfonts{%
	\loadfonts{twenty}{20}%
	\global\def\twentyfonts{}}
\def\twentyfourfonts{%
	\loadfonts{twentyfour}{24}%
	\global\def\twentyfourfonts{}}
\def\famset#1#2#3#4#5{%
	\textfont#1\csname#3#2\endcsname
	\scriptfont#1\csname#4#2\endcsname
	\scriptscriptfont#1\csname#5#2\endcsname}
\def\ninepoint{%
	\ifplain\else\ninefonts\sevenfonts\fivefonts\fi
	\famset0{rm}{nine}{seven}{five}%
	\famset1{i} {nine}{seven}{five}%
	\famset2{sy}{nine}{seven}{five}%
	\famset3{ex}{nine}{seven}{five}%
	\famset\itfam{it}{nine}{seven}{five}%
	\famset\slfam{sl}{nine}{seven}{five}%
	\famset\ttfam{tt}{ten}{seven}{five}%
	\famset\bffam{bf}{nine}{seven}{five}%
	\def\rm{\fam0\ninerm}%
	\def\it{\fam\itfam\nineit}%
	\def\sl{\fam\slfam\ninesl}%
	\def\tt{\fam\ttfam\tentt}%
	\def\bf{\famset0{bf}{nine}{seven}{five}%
		\famset1{mib}{nine}{seven}{five}%
		\famset2{bsy}{nine}{seven}{five}%
		\fam\bffam\ninebf}%
	\setbox\strutbox=\hbox{\vrule height 8pt depth 3pt width 0pt}%
	\baselineskip11pt\rm%
	}
\def\tenpoint{%
	\ifplain\else\tenfonts\sevenfonts\fivefonts\fi
	\famset0{rm}{ten}{seven}{five}%
	\famset1{i} {ten}{seven}{five}%
	\famset2{sy}{ten}{seven}{five}%
	\famset3{ex}{ten}{seven}{five}%
	\famset\itfam{it}{ten}{seven}{five}%
	\famset\slfam{sl}{ten}{seven}{five}%
	\famset\ttfam{tt}{ten}{seven}{five}%
	\famset\bffam{bf}{ten}{seven}{five}%
	\def\rm{\fam0\tenrm}%
	\def\it{\fam\itfam\tenit}%
	\def\sl{\fam\slfam\tensl}%
	\def\tt{\fam\ttfam\tentt}%
	\def\bf{\famset0{bf}{ten}{seven}{five}%
		\famset1{mib}{ten}{seven}{five}%
		\famset2{bsy}{ten}{seven}{five}%
		\fam\bffam\tenbf}%
	\setbox\strutbox=\hbox{\vrule height 8.5pt depth 3.5pt width 0pt}%
	\baselineskip12pt\rm%
	}
\def\twelvepoint{%
	\twelvefonts\ifplain\else\ninefonts\sevenfonts\fi
	\famset0{rm}{twelve}{nine}{seven}%
	\famset1{i} {twelve}{nine}{seven}%
	\famset2{sy}{twelve}{nine}{seven}%
	\famset3{ex}{twelve}{nine}{seven}%
	\famset\itfam{it}{twelve}{nine}{seven}%
	\famset\slfam{sl}{twelve}{nine}{seven}%
	\famset\ttfam{tt}{twelve}{nine}{seven}%
	\famset\bffam{bf}{twelve}{nine}{seven}%
	\def\rm{\fam0\twelverm}%
	\def\it{\fam\itfam\twelveit}%
	\def\sl{\fam\slfam\twelvesl}%
	\def\tt{\fam\ttfam\twelvett}%
	\def\bf{\famset0{bf}{twelve}{nine}{seven}%
		\famset1{mib}{twelve}{nine}{seven}%
		\famset2{bsy}{twelve}{nine}{seven}%
		\fam\bffam\twelvebf}%
	\setbox\strutbox=\hbox{\vrule height 10pt depth 4pt width 0pt}%
	\baselineskip14pt\rm%
	}
\def\fourteenpoint{%
	\fourteenfonts\twelvefonts\ifplain\else\tenfonts\fi
	\famset0{rm}{fourteen}{twelve}{ten}%
	\famset1{i} {fourteen}{twelve}{ten}%
	\famset2{sy}{fourteen}{twelve}{ten}%
	\famset3{ex}{fourteen}{twelve}{ten}%
	\famset\itfam{it}{fourteen}{twelve}{ten}%
	\famset\slfam{sl}{fourteen}{twelve}{ten}%
	\famset\ttfam{tt}{fourteen}{twelve}{ten}%
	\famset\bffam{bf}{fourteen}{twelve}{ten}%
	\def\rm{\fam0\fourteenrm}%
	\def\it{\fam\itfam\fourteenit}%
	\def\sl{\fam\slfam\fourteensl}%
	\def\tt{\fam\ttfam\fourteentt}%
	\def\bf{\famset0{bf}{fourteen}{twelve}{ten}%
		\famset1{mib}{fourteen}{twelve}{ten}%
		\famset2{bsy}{fourteen}{twelve}{ten}%
		\fam\bffam\fourteenbf}%
	\setbox\strutbox=\hbox{\vrule height 12pt depth 5pt width 0pt}%
	\baselineskip17pt\rm%
	}
\def\sixteenpoint{%
	\sixteenfonts\fourteenfonts\twelvefonts
	\famset0{rm}{sixteen}{fourteen}{twelve}%
	\famset1{i} {sixteen}{fourteen}{twelve}%
	\famset2{sy}{sixteen}{fourteen}{twelve}%
	\famset3{ex}{sixteen}{fourteen}{twelve}%
	\famset\itfam{it}{sixteen}{fourteen}{twelve}%
	\famset\slfam{sl}{sixteen}{fourteen}{twelve}%
	\famset\ttfam{tt}{sixteen}{fourteen}{twelve}%
	\famset\bffam{bf}{sixteen}{fourteen}{twelve}%
	\def\rm{\fam0\sixteenrm}%
	\def\it{\fam\itfam\sixteenit}%
	\def\sl{\fam\slfam\sixteensl}%
	\def\tt{\fam\ttfam\sixteentt}%
	\def\bf{\famset0{bf}{sixteen}{fourteen}{twelve}%
		\famset1{mib}{sixteen}{fourteen}{twelve}%
		\famset2{bsy}{sixteen}{fourteen}{twelve}%
		\fam\bffam\sixteenbf}%
	\setbox\strutbox=\hbox{\vrule height 14pt depth 6pt width 0pt}%
	\baselineskip20pt\rm%
	}
\def\twentypoint{%
	\twentyfonts\sixteenfonts\fourteenfonts
	\famset0{rm}{twenty}{sixteen}{fourteen}%
	\famset1{i} {twenty}{sixteen}{fourteen}%
	\famset2{sy}{twenty}{sixteen}{fourteen}%
	\famset3{ex}{twenty}{sixteen}{fourteen}%
	\famset\itfam{it}{twenty}{sixteen}{fourteen}%
	\famset\slfam{sl}{twenty}{sixteen}{fourteen}%
	\famset\ttfam{tt}{twenty}{sixteen}{fourteen}%
	\famset\bffam{bf}{twenty}{sixteen}{fourteen}%
	\def\rm{\fam0\twentyrm}%
	\def\it{\fam\itfam\twentyit}%
	\def\sl{\fam\slfam\twentysl}%
	\def\tt{\fam\ttfam\twentytt}%
	\def\bf{\famset0{bf}{twenty}{sixteen}{fourteen}%
		\famset1{mib}{twenty}{sixteen}{fourteen}%
		\famset2{bsy}{twenty}{sixteen}{fourteen}%
		\fam\bffam\twentybf}%
	\setbox\strutbox=\hbox{\vrule height 17pt depth 7pt width 0pt}%
	\baselineskip24pt\rm%
	}
\def\twentyfourpoint{%
	\twentyfourfonts\twentyfonts\sixteenfonts
	\famset0{rm}{twentyfour}{twenty}{sixteen}%
	\famset1{i} {twentyfour}{twenty}{sixteen}%
	\famset2{sy}{twentyfour}{twenty}{sixteen}%
	\famset3{ex}{twentyfour}{twenty}{sixteen}%
	\famset\itfam{it}{twentyfour}{twenty}{sixteen}%
	\famset\slfam{sl}{twentyfour}{twenty}{sixteen}%
	\famset\ttfam{tt}{twentyfour}{twenty}{sixteen}%
	\famset\bffam{bf}{twentyfour}{twenty}{sixteen}%
	\def\rm{\fam0\twentyfourrm}%
	\def\it{\fam\itfam\twentyfourit}%
	\def\sl{\fam\slfam\twentyfoursl}%
	\def\tt{\fam\ttfam\twentyfourtt}%
	\def\bf{\famset0{bf}{twentyfour}{twenty}{sixteen}%
		\famset1{mib}{twentyfour}{twenty}{sixteen}%
		\famset2{bsy}{twentyfour}{twenty}{sixteen}%
		\fam\bffam\twentyfourbf}%
	\setbox\strutbox=\hbox{\vrule height 17pt depth 7pt width 0pt}%
	\baselineskip24pt\rm%
	}
\expandafter\def\csname fonts are loaded\endcsname{}%

\newif\ifintexdraw
\ifx\axisscale\donotdefinethis\def\axisscale{1}\fi
\def\e{\ifintexdraw\immediate\message{Ending figure}%
       \esegment\etexdraw\ifvmode\medskip\fi\intexdrawfalse\else
       \immediate\message{\string\e\space ignored}\fi}
\def\m#1#2{\move (#1 #2)}
\def\w#1{\linewd #1 }
\def\p#1#2{\move (#1 #2) \fcir f:0 r:1}
\def\l#1#2#3#4{\move (#1 #2) \lvec (#3 #4)}
\def\n#1#2{\lvec (#1 #2)}
\def\s#1#2#3#4{\ifintexdraw\immediate\message{\string\s\space ignored}%
               \else\intexdrawtrue
	       \immediate\message{Starting figure}
               \btexdraw
               \drawdim bp %
               \setunitscale 0.08791 %
	       \bsegment
	       \expandafter\expandafter\expandafter
               \relsegscale\expandafter\axisscale\space
               \fi}
\def\TX{\ifmmode\else$\fi\rm}
\def\XT{\ifmmode$\relax\else\fi}
{\catcode`p=12 \catcode`t=12
 \gdef\dimno#1pt{#1}}%
\def\scle#1#2{{{\dimen0=1000pt
                \dimen0=#2\dimen0\relax
                \expandafter\dimen\expandafter0\expandafter=
                \axisscale\dimen0\relax\divide\dimen0 by 1000\relax
                \xdef\myowntemp{\expandafter\dimno\the\dimen0}}
               \setbox0\hbox{\rotsclTeX{0}{\myowntemp}{\myowntemp}{\hbox{#1}}}%
               \ht0=\myowntemp\ht0
	       \dp0=\myowntemp\dp0
	       \wd0=\myowntemp\wd0
               \box0}}
\def\t#1{\textref h:L v:B \htext{$\rm #1$}}
\def\ltx(#1 #2) #3#4{\textref h:L v:C \htext (#1 #2){\scle{#3\XT}{#4}}}
\def\rtx(#1 #2) #3#4{\textref h:R v:C \htext (#1 #2){\scle{#3\XT}{#4}}}
\def\ctx(#1 #2) #3#4{\textref h:C v:C \htext (#1 #2){\scle{#3\XT}{#4}}}
\def\vtx(#1 #2) #3#4{\textref h:C v:T \vtext (#1 #2){\scle{#3\XT}{#4}}}
\def\solid{\lpatt ()}
\def\disconnected{\lpatt (0 100)}
\def\dotted{\lpatt (10 30)}
\def\dotdashed{\lpatt (10 40 100 100)}
\def\shortdashed{\lpatt (100 100)}
\def\longdashed{\lpatt (200 100)}
\def\dashdotdotted{\lpatt (100 100 10 40 10 40)}
\def\f#1{%
\csname#1\endcsname
}
\let\axisloaded=\relax

\hskip -40pt
\hbox{
\s {0}{0}{4096}{4096}
\f {solid}
\w {15.000000}
\m {750}{551}
\n {801}{551}
\n {851}{551}
\n {902}{551}
\n {952}{551}
\n {1003}{576}
\n {1053}{576}
\n {1104}{576}
\n {1154}{576}
\n {1205}{601}
\n {1256}{601}
\n {1306}{601}
\n {1357}{601}
\n {1407}{626}
\n {1458}{626}
\n {1508}{626}
\n {1559}{652}
\n {1609}{652}
\n {1660}{677}
\n {1711}{677}
\n {1761}{677}
\n {1812}{702}
\n {1862}{702}
\n {1913}{728}
\n {1963}{728}
\n {2014}{753}
\n {2064}{778}
\n {2115}{778}
\n {2166}{803}
\n {2216}{829}
\n {2267}{829}
\n {2317}{854}
\n {2368}{879}
\n {2418}{879}
\n {2469}{904}
\n {2519}{930}
\n {2570}{955}
\n {2621}{980}
\n {2671}{1006}
\n {2722}{1031}
\n {2772}{1056}
\n {2823}{1081}
\n {2873}{1107}
\n {2924}{1132}
\n {2974}{1157}
\n {3025}{1208}
\n {3076}{1233}
\n {3126}{1258}
\n {3177}{1309}
\n {3227}{1334}
\n {3278}{1385}
\n {3328}{1410}
\n {3379}{1461}
\n {3429}{1486}
\n {3480}{1536}
\n {3531}{1587}
\n {3581}{1638}
\n {3632}{1688}
\n {3682}{1739}
\n {3733}{1789}
\n {3783}{1840}
\n {3834}{1890}
\n {3884}{1941}
\n {3935}{2017}
\n {3986}{2067}
\n {4036}{2143}
\n {4087}{2194}
\n {4137}{2269}
\n {4188}{2320}
\n {4238}{2396}
\n {4289}{2472}
\n {4339}{2548}
\n {4390}{2623}
\n {4441}{2724}
\n {4491}{2800}
\n {4542}{2876}
\n {4592}{2977}
\n {4643}{3078}
\n {4693}{3154}
\n {4744}{3255}
\n {4794}{3356}
\n {4845}{3458}
\n {4896}{3584}
\n {4946}{3685}
\n {4997}{3786}
\n {5047}{3913}
\n {5098}{4039}
\n {5148}{4165}
\n {5199}{4292}
\n {5249}{4418}
\n {5300}{4544}
\f {solid}
\m {750}{500}
\n {5300}{500}
\n {5300}{5050}
\n {750}{5050}
\n {750}{500}
\m {851}{500}
\n {851}{550}
\m {851}{5000}
\n {851}{5050}
\m {952}{500}
\n {952}{550}
\m {952}{5000}
\n {952}{5050}
\m {1053}{500}
\n {1053}{550}
\m {1053}{5000}
\n {1053}{5050}
\m {1154}{500}
\n {1154}{550}
\m {1154}{5000}
\n {1154}{5050}
\m {1256}{500}
\n {1256}{600}
\m {1256}{4950}
\n {1256}{5050}
\m {1357}{500}
\n {1357}{550}
\m {1357}{5000}
\n {1357}{5050}
\m {1458}{500}
\n {1458}{550}
\m {1458}{5000}
\n {1458}{5050}
\m {1559}{500}
\n {1559}{550}
\m {1559}{5000}
\n {1559}{5050}
\m {1660}{500}
\n {1660}{550}
\m {1660}{5000}
\n {1660}{5050}
\m {1761}{500}
\n {1761}{600}
\m {1761}{4950}
\n {1761}{5050}
\m {1862}{500}
\n {1862}{550}
\m {1862}{5000}
\n {1862}{5050}
\m {1963}{500}
\n {1963}{550}
\m {1963}{5000}
\n {1963}{5050}
\m {2064}{500}
\n {2064}{550}
\m {2064}{5000}
\n {2064}{5050}
\m {2166}{500}
\n {2166}{550}
\m {2166}{5000}
\n {2166}{5050}
\m {2267}{500}
\n {2267}{600}
\m {2267}{4950}
\n {2267}{5050}
\m {2368}{500}
\n {2368}{550}
\m {2368}{5000}
\n {2368}{5050}
\m {2469}{500}
\n {2469}{550}
\m {2469}{5000}
\n {2469}{5050}
\m {2570}{500}
\n {2570}{550}
\m {2570}{5000}
\n {2570}{5050}
\m {2671}{500}
\n {2671}{550}
\m {2671}{5000}
\n {2671}{5050}
\m {2772}{500}
\n {2772}{600}
\m {2772}{4950}
\n {2772}{5050}
\m {2873}{500}
\n {2873}{550}
\m {2873}{5000}
\n {2873}{5050}
\m {2974}{500}
\n {2974}{550}
\m {2974}{5000}
\n {2974}{5050}
\m {3076}{500}
\n {3076}{550}
\m {3076}{5000}
\n {3076}{5050}
\m {3177}{500}
\n {3177}{550}
\m {3177}{5000}
\n {3177}{5050}
\m {3278}{500}
\n {3278}{600}
\m {3278}{4950}
\n {3278}{5050}
\m {3379}{500}
\n {3379}{550}
\m {3379}{5000}
\n {3379}{5050}
\m {3480}{500}
\n {3480}{550}
\m {3480}{5000}
\n {3480}{5050}
\m {3581}{500}
\n {3581}{550}
\m {3581}{5000}
\n {3581}{5050}
\m {3682}{500}
\n {3682}{550}
\m {3682}{5000}
\n {3682}{5050}
\m {3783}{500}
\n {3783}{600}
\m {3783}{4950}
\n {3783}{5050}
\m {3884}{500}
\n {3884}{550}
\m {3884}{5000}
\n {3884}{5050}
\m {3986}{500}
\n {3986}{550}
\m {3986}{5000}
\n {3986}{5050}
\m {4087}{500}
\n {4087}{550}
\m {4087}{5000}
\n {4087}{5050}
\m {4188}{500}
\n {4188}{550}
\m {4188}{5000}
\n {4188}{5050}
\m {4289}{500}
\n {4289}{600}
\m {4289}{4950}
\n {4289}{5050}
\m {4390}{500}
\n {4390}{550}
\m {4390}{5000}
\n {4390}{5050}
\m {4491}{500}
\n {4491}{550}
\m {4491}{5000}
\n {4491}{5050}
\m {4592}{500}
\n {4592}{550}
\m {4592}{5000}
\n {4592}{5050}
\m {4693}{500}
\n {4693}{550}
\m {4693}{5000}
\n {4693}{5050}
\m {4794}{500}
\n {4794}{600}
\m {4794}{4950}
\n {4794}{5050}
\m {4896}{500}
\n {4896}{550}
\m {4896}{5000}
\n {4896}{5050}
\m {4997}{500}
\n {4997}{550}
\m {4997}{5000}
\n {4997}{5050}
\m {5098}{500}
\n {5098}{550}
\m {5098}{5000}
\n {5098}{5050}
\m {5199}{500}
\n {5199}{550}
\m {5199}{5000}
\n {5199}{5050}
\m {750}{753}
\n {800}{753}
\m {5250}{753}
\n {5300}{753}
\m {750}{1006}
\n {800}{1006}
\m {5250}{1006}
\n {5300}{1006}
\m {750}{1258}
\n {800}{1258}
\m {5250}{1258}
\n {5300}{1258}
\m {750}{1511}
\n {800}{1511}
\m {5250}{1511}
\n {5300}{1511}
\m {750}{1764}
\n {850}{1764}
\m {5200}{1764}
\n {5300}{1764}
\m {750}{2017}
\n {800}{2017}
\m {5250}{2017}
\n {5300}{2017}
\m {750}{2269}
\n {800}{2269}
\m {5250}{2269}
\n {5300}{2269}
\m {750}{2522}
\n {800}{2522}
\m {5250}{2522}
\n {5300}{2522}
\m {750}{2775}
\n {800}{2775}
\m {5250}{2775}
\n {5300}{2775}
\m {750}{3028}
\n {850}{3028}
\m {5200}{3028}
\n {5300}{3028}
\m {750}{3281}
\n {800}{3281}
\m {5250}{3281}
\n {5300}{3281}
\m {750}{3533}
\n {800}{3533}
\m {5250}{3533}
\n {5300}{3533}
\m {750}{3786}
\n {800}{3786}
\m {5250}{3786}
\n {5300}{3786}
\m {750}{4039}
\n {800}{4039}
\m {5250}{4039}
\n {5300}{4039}
\m {750}{4292}
\n {850}{4292}
\m {5200}{4292}
\n {5300}{4292}
\m {750}{4544}
\n {800}{4544}
\m {5250}{4544}
\n {5300}{4544}
\m {750}{4797}
\n {800}{4797}
\m {5250}{4797}
\n {5300}{4797}
\ctx(2729 -2230) {\twentypoint Figure 1} {1.000000}
\ctx(750 320) {\sixteenpoint \TX 1} {0.937500}
\ctx(1256 320) {\sixteenpoint \TX 2} {0.937500}
\ctx(1761 320) {\sixteenpoint \TX 3} {0.937500}
\ctx(2267 320) {\sixteenpoint \TX 4} {0.937500}
\ctx(2772 320) {\sixteenpoint \TX 5} {0.937500}
\ctx(3278 320) {\sixteenpoint \TX 6} {0.937500}
\ctx(3783 320) {\sixteenpoint \TX 7} {0.937500}
\ctx(4289 320) {\sixteenpoint \TX 8} {0.937500}
\ctx(4794 320) {\sixteenpoint \TX 9} {0.937500}
\ctx(5300 320) {\sixteenpoint \TX 10} {0.937500}
\rtx(675 500) {\sixteenpoint \TX 0} {0.937500}
\rtx(675 1764) {\sixteenpoint \TX 50} {0.937500}
\rtx(675 3028) {\sixteenpoint \TX 100} {0.937500}
\rtx(675 4292) {\sixteenpoint \TX 150} {0.937500}
\ctx(3025 140) {\sixteenpoint \TX -\log_{10}{\bar x}} {0.937500}
\vtx(-75 2775) {\sixteenpoint \TX n_{\rm min}} {0.937500}
\e 
}
\vfill\eject

\hskip -40 pt
\hbox{
\s {0}{0}{4096}{4096}
\f {solid}
\w {15.000000}
\m {750}{4951}
\n {801}{4944}
\n {851}{4939}
\n {902}{4935}
\n {952}{4931}
\n {1003}{4933}
\n {1053}{4928}
\n {1104}{4923}
\n {1154}{4918}
\n {1205}{4918}
\n {1256}{4912}
\n {1306}{4906}
\n {1357}{4901}
\n {1407}{4898}
\n {1458}{4891}
\n {1508}{4885}
\n {1559}{4880}
\n {1609}{4873}
\n {1660}{4868}
\n {1711}{4859}
\n {1761}{4850}
\n {1812}{4843}
\n {1862}{4833}
\n {1913}{4825}
\n {1963}{4814}
\n {2014}{4805}
\n {2064}{4794}
\n {2115}{4782}
\n {2166}{4770}
\n {2216}{4758}
\n {2267}{4743}
\n {2317}{4729}
\n {2368}{4714}
\n {2418}{4697}
\n {2469}{4680}
\n {2519}{4663}
\n {2570}{4644}
\n {2621}{4625}
\n {2671}{4604}
\n {2722}{4582}
\n {2772}{4560}
\n {2823}{4536}
\n {2873}{4511}
\n {2924}{4485}
\n {2974}{4457}
\n {3025}{4429}
\n {3076}{4399}
\n {3126}{4367}
\n {3177}{4335}
\n {3227}{4300}
\n {3278}{4264}
\n {3328}{4226}
\n {3379}{4188}
\n {3429}{4146}
\n {3480}{4104}
\n {3531}{4059}
\n {3581}{4013}
\n {3632}{3965}
\n {3682}{3914}
\n {3733}{3861}
\n {3783}{3807}
\n {3834}{3749}
\n {3884}{3690}
\n {3935}{3628}
\n {3986}{3564}
\n {4036}{3497}
\n {4087}{3427}
\n {4137}{3355}
\n {4188}{3279}
\n {4238}{3201}
\n {4289}{3120}
\n {4339}{3035}
\n {4390}{2948}
\n {4441}{2857}
\n {4491}{2762}
\n {4542}{2664}
\n {4592}{2563}
\n {4643}{2457}
\n {4693}{2348}
\n {4744}{2234}
\n {4794}{2117}
\n {4845}{1995}
\n {4896}{1868}
\n {4946}{1737}
\n {4997}{1602}
\n {5047}{1461}
\n {5098}{1316}
\n {5148}{1165}
\n {5199}{1010}
\n {5249}{848}
\n {5300}{681}
\f {solid}
\m {750}{500}
\n {5300}{500}
\n {5300}{5050}
\n {750}{5050}
\n {750}{500}
\m {851}{500}
\n {851}{550}
\m {851}{5000}
\n {851}{5050}
\m {952}{500}
\n {952}{550}
\m {952}{5000}
\n {952}{5050}
\m {1053}{500}
\n {1053}{550}
\m {1053}{5000}
\n {1053}{5050}
\m {1154}{500}
\n {1154}{550}
\m {1154}{5000}
\n {1154}{5050}
\m {1256}{500}
\n {1256}{600}
\m {1256}{4950}
\n {1256}{5050}
\m {1357}{500}
\n {1357}{550}
\m {1357}{5000}
\n {1357}{5050}
\m {1458}{500}
\n {1458}{550}
\m {1458}{5000}
\n {1458}{5050}
\m {1559}{500}
\n {1559}{550}
\m {1559}{5000}
\n {1559}{5050}
\m {1660}{500}
\n {1660}{550}
\m {1660}{5000}
\n {1660}{5050}
\m {1761}{500}
\n {1761}{600}
\m {1761}{4950}
\n {1761}{5050}
\m {1862}{500}
\n {1862}{550}
\m {1862}{5000}
\n {1862}{5050}
\m {1963}{500}
\n {1963}{550}
\m {1963}{5000}
\n {1963}{5050}
\m {2064}{500}
\n {2064}{550}
\m {2064}{5000}
\n {2064}{5050}
\m {2166}{500}
\n {2166}{550}
\m {2166}{5000}
\n {2166}{5050}
\m {2267}{500}
\n {2267}{600}
\m {2267}{4950}
\n {2267}{5050}
\m {2368}{500}
\n {2368}{550}
\m {2368}{5000}
\n {2368}{5050}
\m {2469}{500}
\n {2469}{550}
\m {2469}{5000}
\n {2469}{5050}
\m {2570}{500}
\n {2570}{550}
\m {2570}{5000}
\n {2570}{5050}
\m {2671}{500}
\n {2671}{550}
\m {2671}{5000}
\n {2671}{5050}
\m {2772}{500}
\n {2772}{600}
\m {2772}{4950}
\n {2772}{5050}
\m {2873}{500}
\n {2873}{550}
\m {2873}{5000}
\n {2873}{5050}
\m {2974}{500}
\n {2974}{550}
\m {2974}{5000}
\n {2974}{5050}
\m {3076}{500}
\n {3076}{550}
\m {3076}{5000}
\n {3076}{5050}
\m {3177}{500}
\n {3177}{550}
\m {3177}{5000}
\n {3177}{5050}
\m {3278}{500}
\n {3278}{600}
\m {3278}{4950}
\n {3278}{5050}
\m {3379}{500}
\n {3379}{550}
\m {3379}{5000}
\n {3379}{5050}
\m {3480}{500}
\n {3480}{550}
\m {3480}{5000}
\n {3480}{5050}
\m {3581}{500}
\n {3581}{550}
\m {3581}{5000}
\n {3581}{5050}
\m {3682}{500}
\n {3682}{550}
\m {3682}{5000}
\n {3682}{5050}
\m {3783}{500}
\n {3783}{600}
\m {3783}{4950}
\n {3783}{5050}
\m {3884}{500}
\n {3884}{550}
\m {3884}{5000}
\n {3884}{5050}
\m {3986}{500}
\n {3986}{550}
\m {3986}{5000}
\n {3986}{5050}
\m {4087}{500}
\n {4087}{550}
\m {4087}{5000}
\n {4087}{5050}
\m {4188}{500}
\n {4188}{550}
\m {4188}{5000}
\n {4188}{5050}
\m {4289}{500}
\n {4289}{600}
\m {4289}{4950}
\n {4289}{5050}
\m {4390}{500}
\n {4390}{550}
\m {4390}{5000}
\n {4390}{5050}
\m {4491}{500}
\n {4491}{550}
\m {4491}{5000}
\n {4491}{5050}
\m {4592}{500}
\n {4592}{550}
\m {4592}{5000}
\n {4592}{5050}
\m {4693}{500}
\n {4693}{550}
\m {4693}{5000}
\n {4693}{5050}
\m {4794}{500}
\n {4794}{600}
\m {4794}{4950}
\n {4794}{5050}
\m {4896}{500}
\n {4896}{550}
\m {4896}{5000}
\n {4896}{5050}
\m {4997}{500}
\n {4997}{550}
\m {4997}{5000}
\n {4997}{5050}
\m {5098}{500}
\n {5098}{550}
\m {5098}{5000}
\n {5098}{5050}
\m {5199}{500}
\n {5199}{550}
\m {5199}{5000}
\n {5199}{5050}
\m {750}{712}
\n {850}{712}
\m {5200}{712}
\n {5300}{712}
\m {750}{923}
\n {800}{923}
\m {5250}{923}
\n {5300}{923}
\m {750}{1135}
\n {800}{1135}
\m {5250}{1135}
\n {5300}{1135}
\m {750}{1347}
\n {800}{1347}
\m {5250}{1347}
\n {5300}{1347}
\m {750}{1558}
\n {800}{1558}
\m {5250}{1558}
\n {5300}{1558}
\m {750}{1770}
\n {850}{1770}
\m {5200}{1770}
\n {5300}{1770}
\m {750}{1981}
\n {800}{1981}
\m {5250}{1981}
\n {5300}{1981}
\m {750}{2193}
\n {800}{2193}
\m {5250}{2193}
\n {5300}{2193}
\m {750}{2405}
\n {800}{2405}
\m {5250}{2405}
\n {5300}{2405}
\m {750}{2616}
\n {800}{2616}
\m {5250}{2616}
\n {5300}{2616}
\m {750}{2828}
\n {850}{2828}
\m {5200}{2828}
\n {5300}{2828}
\m {750}{3040}
\n {800}{3040}
\m {5250}{3040}
\n {5300}{3040}
\m {750}{3251}
\n {800}{3251}
\m {5250}{3251}
\n {5300}{3251}
\m {750}{3463}
\n {800}{3463}
\m {5250}{3463}
\n {5300}{3463}
\m {750}{3674}
\n {800}{3674}
\m {5250}{3674}
\n {5300}{3674}
\m {750}{3886}
\n {850}{3886}
\m {5200}{3886}
\n {5300}{3886}
\m {750}{4098}
\n {800}{4098}
\m {5250}{4098}
\n {5300}{4098}
\m {750}{4309}
\n {800}{4309}
\m {5250}{4309}
\n {5300}{4309}
\m {750}{4521}
\n {800}{4521}
\m {5250}{4521}
\n {5300}{4521}
\m {750}{4733}
\n {800}{4733}
\m {5250}{4733}
\n {5300}{4733}
\m {750}{4944}
\n {850}{4944}
\m {5200}{4944}
\n {5300}{4944}
\ctx(2661 -2230) {\twentypoint Figure 2} {1.000000}
\ctx(750 320) {\sixteenpoint \TX 1} {0.937500}
\ctx(1256 320) {\sixteenpoint \TX 2} {0.937500}
\ctx(1761 320) {\sixteenpoint \TX 3} {0.937500}
\ctx(2267 320) {\sixteenpoint \TX 4} {0.937500}
\ctx(2772 320) {\sixteenpoint \TX 5} {0.937500}
\ctx(3278 320) {\sixteenpoint \TX 6} {0.937500}
\ctx(3783 320) {\sixteenpoint \TX 7} {0.937500}
\ctx(4289 320) {\sixteenpoint \TX 8} {0.937500}
\ctx(4794 320) {\sixteenpoint \TX 9} {0.937500}
\ctx(5300 320) {\sixteenpoint \TX 10} {0.937500}
\rtx(675 712) {\sixteenpoint \TX -400} {0.937500}
\rtx(675 1770) {\sixteenpoint \TX -300} {0.937500}
\rtx(675 2828) {\sixteenpoint \TX -200} {0.937500}
\rtx(675 3886) {\sixteenpoint \TX -100} {0.937500}
\rtx(675 4944) {\sixteenpoint \TX 0} {0.937500}
\ctx(3025 140) {\sixteenpoint \TX -\log_{10}{\bar x}} {0.937500}
\vtx(-165 2775) {\sixteenpoint \TX \log_{10}{{R^\ast}\over x}} {0.937500}
\e 
}
\vfill\eject

\hskip -40pt
\hbox{
\s {0}{0}{4096}{4096}
\f {solid}
\w {15.000000}
\m {750}{4681}
\n {801}{3304}
\n {851}{2518}
\n {902}{1927}
\n {952}{1433}
\n {1003}{2590}
\n {1053}{2077}
\n {1104}{1628}
\n {1154}{1220}
\n {1205}{2043}
\n {1256}{1632}
\n {1306}{1249}
\n {1357}{886}
\n {1407}{1505}
\n {1458}{1144}
\n {1508}{796}
\n {1559}{1290}
\n {1609}{945}
\n {1660}{1351}
\n {1711}{1010}
\n {1761}{678}
\n {1812}{1010}
\n {1862}{681}
\n {1913}{957}
\n {1963}{633}
\n {2014}{862}
\n {2064}{1054}
\n {2115}{732}
\n {2166}{890}
\n {2216}{1021}
\n {2267}{702}
\n {2317}{807}
\n {2368}{891}
\n {2418}{575}
\n {2469}{639}
\n {2519}{687}
\n {2570}{718}
\n {2621}{735}
\n {2671}{739}
\n {2722}{731}
\n {2772}{712}
\n {2823}{684}
\n {2873}{646}
\n {2924}{600}
\n {2974}{546}
\n {3025}{724}
\n {3076}{649}
\n {3126}{567}
\n {3177}{695}
\n {3227}{597}
\n {3278}{694}
\n {3328}{582}
\n {3379}{652}
\n {3429}{528}
\n {3480}{576}
\n {3531}{610}
\n {3581}{631}
\n {3632}{641}
\n {3682}{640}
\n {3733}{630}
\n {3783}{609}
\n {3834}{579}
\n {3884}{546}
\n {3935}{627}
\n {3986}{573}
\n {4036}{630}
\n {4087}{561}
\n {4137}{597}
\n {4188}{512}
\n {4238}{530}
\n {4289}{536}
\n {4339}{530}
\n {4390}{518}
\n {4441}{591}
\n {4491}{561}
\n {4542}{521}
\n {4592}{558}
\n {4643}{588}
\n {4693}{524}
\n {4744}{533}
\n {4794}{530}
\n {4845}{524}
\n {4896}{573}
\n {4946}{546}
\n {4997}{512}
\n {5047}{536}
\n {5098}{546}
\n {5148}{552}
\n {5199}{546}
\n {5249}{530}
\n {5300}{509}
\f {solid}
\m {750}{500}
\n {5300}{500}
\n {5300}{5050}
\n {750}{5050}
\n {750}{500}
\m {851}{500}
\n {851}{550}
\m {851}{5000}
\n {851}{5050}
\m {952}{500}
\n {952}{550}
\m {952}{5000}
\n {952}{5050}
\m {1053}{500}
\n {1053}{550}
\m {1053}{5000}
\n {1053}{5050}
\m {1154}{500}
\n {1154}{550}
\m {1154}{5000}
\n {1154}{5050}
\m {1256}{500}
\n {1256}{600}
\m {1256}{4950}
\n {1256}{5050}
\m {1357}{500}
\n {1357}{550}
\m {1357}{5000}
\n {1357}{5050}
\m {1458}{500}
\n {1458}{550}
\m {1458}{5000}
\n {1458}{5050}
\m {1559}{500}
\n {1559}{550}
\m {1559}{5000}
\n {1559}{5050}
\m {1660}{500}
\n {1660}{550}
\m {1660}{5000}
\n {1660}{5050}
\m {1761}{500}
\n {1761}{600}
\m {1761}{4950}
\n {1761}{5050}
\m {1862}{500}
\n {1862}{550}
\m {1862}{5000}
\n {1862}{5050}
\m {1963}{500}
\n {1963}{550}
\m {1963}{5000}
\n {1963}{5050}
\m {2064}{500}
\n {2064}{550}
\m {2064}{5000}
\n {2064}{5050}
\m {2166}{500}
\n {2166}{550}
\m {2166}{5000}
\n {2166}{5050}
\m {2267}{500}
\n {2267}{600}
\m {2267}{4950}
\n {2267}{5050}
\m {2368}{500}
\n {2368}{550}
\m {2368}{5000}
\n {2368}{5050}
\m {2469}{500}
\n {2469}{550}
\m {2469}{5000}
\n {2469}{5050}
\m {2570}{500}
\n {2570}{550}
\m {2570}{5000}
\n {2570}{5050}
\m {2671}{500}
\n {2671}{550}
\m {2671}{5000}
\n {2671}{5050}
\m {2772}{500}
\n {2772}{600}
\m {2772}{4950}
\n {2772}{5050}
\m {2873}{500}
\n {2873}{550}
\m {2873}{5000}
\n {2873}{5050}
\m {2974}{500}
\n {2974}{550}
\m {2974}{5000}
\n {2974}{5050}
\m {3076}{500}
\n {3076}{550}
\m {3076}{5000}
\n {3076}{5050}
\m {3177}{500}
\n {3177}{550}
\m {3177}{5000}
\n {3177}{5050}
\m {3278}{500}
\n {3278}{600}
\m {3278}{4950}
\n {3278}{5050}
\m {3379}{500}
\n {3379}{550}
\m {3379}{5000}
\n {3379}{5050}
\m {3480}{500}
\n {3480}{550}
\m {3480}{5000}
\n {3480}{5050}
\m {3581}{500}
\n {3581}{550}
\m {3581}{5000}
\n {3581}{5050}
\m {3682}{500}
\n {3682}{550}
\m {3682}{5000}
\n {3682}{5050}
\m {3783}{500}
\n {3783}{600}
\m {3783}{4950}
\n {3783}{5050}
\m {3884}{500}
\n {3884}{550}
\m {3884}{5000}
\n {3884}{5050}
\m {3986}{500}
\n {3986}{550}
\m {3986}{5000}
\n {3986}{5050}
\m {4087}{500}
\n {4087}{550}
\m {4087}{5000}
\n {4087}{5050}
\m {4188}{500}
\n {4188}{550}
\m {4188}{5000}
\n {4188}{5050}
\m {4289}{500}
\n {4289}{600}
\m {4289}{4950}
\n {4289}{5050}
\m {4390}{500}
\n {4390}{550}
\m {4390}{5000}
\n {4390}{5050}
\m {4491}{500}
\n {4491}{550}
\m {4491}{5000}
\n {4491}{5050}
\m {4592}{500}
\n {4592}{550}
\m {4592}{5000}
\n {4592}{5050}
\m {4693}{500}
\n {4693}{550}
\m {4693}{5000}
\n {4693}{5050}
\m {4794}{500}
\n {4794}{600}
\m {4794}{4950}
\n {4794}{5050}
\m {4896}{500}
\n {4896}{550}
\m {4896}{5000}
\n {4896}{5050}
\m {4997}{500}
\n {4997}{550}
\m {4997}{5000}
\n {4997}{5050}
\m {5098}{500}
\n {5098}{550}
\m {5098}{5000}
\n {5098}{5050}
\m {5199}{500}
\n {5199}{550}
\m {5199}{5000}
\n {5199}{5050}
\m {750}{803}
\n {800}{803}
\m {5250}{803}
\n {5300}{803}
\m {750}{1107}
\n {800}{1107}
\m {5250}{1107}
\n {5300}{1107}
\m {750}{1410}
\n {800}{1410}
\m {5250}{1410}
\n {5300}{1410}
\m {750}{1713}
\n {800}{1713}
\m {5250}{1713}
\n {5300}{1713}
\m {750}{2017}
\n {850}{2017}
\m {5200}{2017}
\n {5300}{2017}
\m {750}{2320}
\n {800}{2320}
\m {5250}{2320}
\n {5300}{2320}
\m {750}{2623}
\n {800}{2623}
\m {5250}{2623}
\n {5300}{2623}
\m {750}{2927}
\n {800}{2927}
\m {5250}{2927}
\n {5300}{2927}
\m {750}{3230}
\n {800}{3230}
\m {5250}{3230}
\n {5300}{3230}
\m {750}{3533}
\n {850}{3533}
\m {5200}{3533}
\n {5300}{3533}
\m {750}{3837}
\n {800}{3837}
\m {5250}{3837}
\n {5300}{3837}
\m {750}{4140}
\n {800}{4140}
\m {5250}{4140}
\n {5300}{4140}
\m {750}{4443}
\n {800}{4443}
\m {5250}{4443}
\n {5300}{4443}
\m {750}{4747}
\n {800}{4747}
\m {5250}{4747}
\n {5300}{4747}
\ctx(2729 -2230) {\twentypoint Figure 3} {1.000000}
\ctx(750 320) {\sixteenpoint \TX 1} {0.937500}
\ctx(1256 320) {\sixteenpoint \TX 2} {0.937500}
\ctx(1761 320) {\sixteenpoint \TX 3} {0.937500}
\ctx(2267 320) {\sixteenpoint \TX 4} {0.937500}
\ctx(2772 320) {\sixteenpoint \TX 5} {0.937500}
\ctx(3278 320) {\sixteenpoint \TX 6} {0.937500}
\ctx(3783 320) {\sixteenpoint \TX 7} {0.937500}
\ctx(4289 320) {\sixteenpoint \TX 8} {0.937500}
\ctx(4794 320) {\sixteenpoint \TX 9} {0.937500}
\ctx(5300 320) {\sixteenpoint \TX 10} {0.937500}
\rtx(675 500) {\sixteenpoint \TX 0} {0.937500}
\rtx(675 2017) {\sixteenpoint \TX 0.5} {0.937500}
\rtx(675 3533) {\sixteenpoint \TX 1} {0.937500}
\rtx(675 5050) {\sixteenpoint \TX 1.5} {0.937500}
\ctx(3025 140) {\sixteenpoint \TX -\log_{10}{\bar x}} {0.937500}
\vtx(-75 2775) {\sixteenpoint \TX \Delta} {0.937500}
\e 
}
\vfill\eject


\begin{references}
\bibitem{df} A.~J.~Dragt and J.~M.~Finn, J.~Math.~Phys. {\bf 17}, 2215 (1976).

\bibitem{fassben}  F.~Fasso and G.~Benettin, J.~Appl.~Math.~Phys. (ZAMP)
{\bf 40}, 307 (1989).

\bibitem{fasso} F.~Fasso, J.~Appl.~Math.~Phys. (ZAMP) {\bf 41}, 843 (1990).

\bibitem{koseleff} P.~V.~Koseleff, {\it Formal Calculus for Lie Methods in 
Hamiltonian Mechanics}, Doctoral Thesis, Ecole Polytechnique, Paris (1993)
(in French).

\bibitem{me} I.~Gjaja, J.~Math.~Phys. {\bf 35}, 1361 (1994).

\bibitem{lt}  See for example G.~E.~O.~Giacaglia,
{\it Perturbation Methods in Non--Linear Systems} (Springer--Verlag, New York,
1972);  
J.~R.~Cary, Phys.~Rep. {\bf 79}, 129 (1981);  
A.~J.~Dragt, in
{\it Physics of High Energy Accelerators}, AIP Conf.~Proc.~ No.84, edited by
R.~A.~Carrigan, F.~R.~Huson, and M.~Month (AIP, New York, 1982) p.147;  
J.~M.~Finn, in {\it Local and Global Methods of Nonlinear Dynamics}, 
Lecture Notes in Physics 252, edited by A.~W.~S\'aenz, W.~W.~Zachary, 
and R.~Cawley (Springer--Verlag, New York 1986) p.63; and
P.~Lochak and
C.~Meunier, {\it Multiphase Averaging for Classical Systems}
(Springer--Verlag, New York, 1988) Appendix 7.

\bibitem{grobner}  W.~Gr\"obner, {\it Die Lie--Reihen und ihre Anwendungen}
(VEB Deutscher Verlag der Wissenschaften, Berlin 1960) (in German) and
 W.~Gr\"obner and F.~Cap, Report, Office of Naval Research
contract N62558-2992 (1962).

\bibitem{riordan} See for example J.~Riordan, {\it An Introduction to
Combinatorial Analysis} (John Wiley, New York, 1958).

\bibitem{bender} See for example C.~M.~Bender and S.~A.~Orszag,
{\it Advanced Mathematical Methods for Scientists and Engineers}
(McGraw--Hill, New York, 1978).

\end{references}
\end{document}